\documentclass[aps,prd,twocolumn,groupedaddress,nofootinbib,longbibliography]{revtex4-1}

\usepackage{amssymb}
\usepackage{amsmath}
\usepackage[linktocpage=true]{hyperref}
\usepackage{graphicx}

\newcommand{\eqast}{\stackrel{*}{=}}
\newcommand{\eqH}{\stackrel{\rm H}{=}}

\begin{document}

\title{On geometry of deformed black holes:\\
       II. Schwarzschild hole surrounded by a Bach--Weyl ring}

\author{M. Basovn\'{\i}k}
\email[]{mbasovnik@gmail.com}
%\homepage[]{Your web page}
%\thanks{}

\author{O. Semer\'ak}
\email[]{oldrich.semerak@mff.cuni.cz}
%\homepage[]{Your web page}
%\thanks{}

\affiliation{Institute of Theoretical Physics, Faculty of Mathematics and Physics,
             Charles University in Prague, Czech Republic}

\date{\today}

\begin{abstract}
We continue to study the response of black-hole space-times on the presence of additional strong sources of gravity. Restricting ourselves to static and axially symmetric (electro-)vacuum exact solutions of Einstein's equations, we first considered the Majumdar--Papapetrou solution for a binary of extreme black holes in a previous paper, while here we deal with a Schwarzschild black hole surrounded by a concentric thin ring described by the Bach--Weyl solution. The geometry is again revealed on the simplest invariants determined by the metric (lapse function) and its gradient (gravitational acceleration), and by curvature (Kretschmann scalar). Extending the metric inside the black hole along null geodesics tangent to the horizon, we mainly focus on the black-hole interior (specifically, on its sections at constant Killing time) where the quantities behave in a way indicating a surprisingly strong influence of the external source. Being already distinct on the level of potential and acceleration, this is still more pronounced on the level of curvature: for a sufficiently massive and/or nearby (small) ring, the Kretschmann scalar even becomes negative in certain toroidal regions mostly touching the horizon from inside. Such regions have been interpreted as those where magnetic-type curvature dominates, but here we deal with space-times which do {\em not} involve rotation and the negative value is achieved due to the {\em electric}-type components of the Riemann/Weyl tensor. The Kretschmann scalar also shapes rather non-trivial landscapes {\em outside} the horizon.
\end{abstract}

\pacs{0420Jb, 0440Nr, 0470Bw}

\maketitle

\section{Introduction}

Interaction of black holes with other gravitating sources is interesting for purely theoretical reasons (non-linear superposition in a strong-field regime) as well as within models of certain astrophysical sources. A black-hole near field is hard to modify significantly as regards potential and intensity, but its higher derivatives (curvature) may be affected by external sources considerably. Here we try to learn and visualize this effect on a Schwarzschild black hole subject to a presence of a concentric static and axially symmetric thin ring described by the Bach--Weyl solution. More specifically, we analyse the behaviour of the simplest invariants given by the metric and its first and second derivatives in dependence on parameters of the system, namely relative mass and radius of the ring. A special attention is given to the black-hole interior, including the vicinity of the central singularity.

In a previous paper \cite{SemerakB-16}, we tried to deform the black-hole field by another black hole and for that purpose we considered the Majumdar--Papapetrou binary system, made of two extremally charged black holes. Though ``the other black hole" is a very strong source, we found that below the horizon the field is not much deformed within that class of space-times. This is connected with the extreme character of their horizons. Indeed, extreme charges are required as sources of the electrostatic field which just compensates the gravitational attraction; otherwise the holes would fall towards each other or would have to be kept static by an even more unphysical strut(s). Therefore, in the present paper we try to distort a black hole which is far from extreme state. Without the electrostatic repulsion, the external source has to be supported by pressure (hoop stresses) or by centrifugal force. The simplest configuration of this kind involves a thin ring or dics surrounding the hole in a static and axially symmetric, concentric manner. Such a setting may capture at least some features of the accreting black holes studied in astrophysics, while still allowing for an exact analytical treatment.

In section \ref{metric-functions}, we first compose the total metric and analyse its behaviour at the horizon. Then in section \ref{below-horizon} we extend the metric to the black-hole interior by solving Einstein's equations numerically along null geodesics starting tangentially to the horizon. In section \ref{invariants}, we compute and visualize on contours the behaviour of the basic invariants in dependence on parameters of the system, namely relative mass of the Bach--Weyl ring and its radius. Some more attention is devoted to the Kretschmann scalar and to the regions where it turns negative, in particular to their relation with the Gauss curvature of the horizon (subsection \ref{Kretschmann}). Final section \ref{concluding} concludes with a summary, a brief scan of similar literature, a remark concerning visualization and some further plans. More details on the null geodesics important for extension of the metric inside the black hole are shifted to Appendix \ref{appendix-A} and the question of extension of the Weyl coordinates is treated in Appendix \ref{appendix-B}. Let us stress that when speaking of ``black hole", we everywhere have in mind a section of the 3D horizon given by constant Killing time ($t$).

Note on notation: equations/values valid on the horizon will be denoted by the index `H', $X\eqH Y$, while expansions valid there will be denoted by an asterisk, $X\eqast Y$.
The black-hole mass is called $M$, while the ring mass ${\cal M}$ and its Weyl radius $b$.
The Weyl-radius coordinate will be denoted by $\rho$; below horizon where it is pure imaginary, we will introduce $\varrho$ by $\rho=:{\rm i}\varrho$.
We use geometrized units in which $c=1$, $G=1$, index-posed comma/semicolon indicates partial/covariant derivative and usual summation rule is employed. Signature of the space-time metric $g_{\mu\nu}$ is ($-$+++), Riemann tensor is defined according to $V_{\nu;\kappa\lambda}-V_{\nu;\lambda\kappa}={R^\mu}_{\nu\kappa\lambda}V_\mu$
and Ricci tensor by $R_{\nu\lambda}={R^\kappa}_{\nu\kappa\lambda}$.
Cosmological constant is set zero.

\section{Weyl metric for Schwarzschild plus ring}
\label{metric-functions}

All vacuum static and axially symmetric space-times can be described by the Weyl-type metric
\begin{equation}  \label{Weyl-metric}
  {\rm d}s^2=-e^{2\nu}{\rm d}t^2+\rho^2 e^{-2\nu}{\rm d}\phi^2
             +e^{2\lambda-2\nu}({\rm d}\rho^2+{\rm d}z^2) \,,
\end{equation}
where $t$ and $\phi$ are Killing time and azimuthal coordinates, and the unknown functions $\nu$ and $\lambda$ depend only on cylindrical-type radius $\rho$ and the ``vertical" linear coordinate $z$ which cover the meridional planes (orthogonal to both Killing directions) in an isotropic manner.
Einstein's equations reduce to
\begin{align}
  &\nu_{,\rho\rho}+\frac{\nu_{,\rho}}{\rho}+\nu_{,zz}=0 \,, \\
  &\lambda_{,\rho}=\rho(\nu_{,\rho})^2-\rho(\nu_{,z})^2 \,, \quad
   \lambda_{,z}=2\rho\,\nu_{,\rho}\nu_{,z} \,,  \label{Einstein-eqs}
\end{align}
i.e. to the Laplace equation and a simple line integral (which is however only rarely solvable explicitely).
Hence, the potential $\nu$ behaves like in Newtonian theory and adds linearly, whereas the second function $\lambda$ does not ``superpose" that simply. For two sources, with $\nu_{1}$ and $\nu_{2}$ denoting their individual potentials, one can write $\lambda=\lambda_{1}+\lambda_{2}+\lambda_{\rm int}$, where $\lambda_{1}$ and $\lambda_{2}$ describe the first and the second source alone (i.e., they satisfy the above equations with just $\nu_{1}$ and $\nu_{2}$, respectively) and $\lambda_{\rm int}$ is the interaction term which is given by
\begin{align}
  \lambda_{{\rm int},\rho}
  &=2\rho\left(\nu_{1,\rho}\nu_{2,\rho}-\nu_{1,z}\nu_{2,z}\right), \\
  \lambda_{{\rm int},z}
  &=2\rho\left(\nu_{1,\rho}\nu_{2,z}+\nu_{1,z}\nu_{2,\rho}\right).
\end{align}
Typically, the potential $\nu$ scales linearly with the source mass, hence $\lambda$ scales with the mass square.

We are specifically interested in space-time generated by a Schwarzschild-type black hole surrounded by a thin ring described by the Bach--Weyl solution.
The Schwarzschild solution appears, respectively in the Weyl and Schwarzschild coordinates, as
\begin{align}
  \nu_{\rm Schw}&= \frac{1}{2}\,\ln\frac{d_1+d_2-2M}{d_1+d_2+2M} \\
                &= \frac{1}{2}\,\ln\left(1-\frac{2M}{r}\right), \\
  \lambda_{\rm Schw}&= \frac{1}{2}\,\ln\frac{(d_1+d_2)^2-4M^2}{4d_1 d_2} \\
                    &= \frac{1}{2}\,\ln\frac{r(r-2M)}{(r-M)^2-M^2\cos^2\theta} \; ,
\end{align}
where
\[d_{1,2}:=\sqrt{\rho^2+(z\mp M)^2}=r-M\mp M\cos\theta \,.\]
Transformation between the coordinates reads
\begin{align}
  \rho=\sqrt{r(r-2M)}\,\sin\theta \,, &\quad
  z=(r-M)\cos\theta \,;  \label{Weyl-Schw} \\
  r-M=\frac{d_2+d_1}{2} \,, &\quad
  M\cos\theta=\frac{d_2-d_1}{2} \,.  \label{Schw-Weyl}
\end{align}
Let us stress that these relations can only be safely used above the horizon (see Appendix \ref{appendix-B}).

It is worth noting that in the case of a Schwarzschild-type centre ($\nu_{1}\equiv\nu_{\rm Schw}$) the field equations for $\lambda$ appear quite simple in Schwarzschild coordinates when expressed in terms of $\lambda_{\rm int}$. Actually, after transforming ($X$ is some quantity)
\begin{align}
  X_{,r} &= X_{,\rho}\rho_{,r}+X_{,z}z_{,r}=  \nonumber \\
         &= X_{,\rho}\;\frac{r-M}{\sqrt{r(r-2M)}}\,\sin\theta+
            X_{,z}\cos\theta \,,  \nonumber \\
  X_{,\theta} &= X_{,\rho}\rho_{,\theta}+X_{,z}z_{,\theta}=  \nonumber \\
              &= X_{,\rho}\,\sqrt{r(r-2M)}\,\cos\theta-
                 X_{,z}(r-M)\sin\theta \,,  \nonumber \\
  \nu_{{\rm Schw},\rho} &= \frac{(d_1+d_2)\,[4M^2-(d_2-d_1)^2]}
                                {8M\rho\;d_1 d_2}  \label{nuSchw,rho} \\
                        &= \frac{M(r-M)\sin\theta}
                                {[(r-M)^2-M^2\cos^2\theta]\,\sqrt{r(r-2M)}} \;, \\
  \nu_{{\rm Schw},z} &= \frac{d_2-d_1}{2\,d_1 d_2}
                      = \frac{M\cos\theta}{(r-M)^2-M^2\cos^2\theta} \;,
                        \label{nuSchw,z}
\end{align}
they lead to
\begin{equation}  \label{lambda_int,eqns}
  \lambda_{{\rm int},r}=\frac{2M\nu_{2,\rho}}{\rho}\,\sin^2\theta \,,
  \quad
  \lambda_{{\rm int},\theta}=-2M\nu_{2,z}\sin\theta \,.
\end{equation}
Therefore, if $\nu_2$ depends linearly on the ``external"-source mass (we will call it ${\cal M}$), then $\lambda_{\rm int}$ is linear in it, too, while $\lambda_2$ is quadratic. Hence, in the decomposition of $\lambda$ the ${\cal M}$ parameter appears as
\begin{equation}  \label{lambda-decomp}
  \lambda=\lambda_{\rm Schw}+\lambda_{\rm int}+\lambda_2
         =\lambda_{\rm Schw}+{\cal M}\,\tilde\lambda_{\rm int}+{\cal M}^2\,\tilde\lambda_2 \,,
\end{equation}
where the pure-Schwarzschild term $\lambda_{\rm Schw}$ as well as the tilded functions $\tilde\lambda_2$ and $\tilde\lambda_{\rm int}$ do not depend on ${\cal M}$.

Our ``second" source is a thin ring with Weyl radius $\rho=b$ and mass ${\cal M}$, described by the Bach--Weyl solution
\begin{align}
  \nu_{\rm BW} &= -\frac{2{\cal M}K(k)}{\pi l_2} \,, \qquad l_{1,2}:=\sqrt{(\rho\mp b)^2+z^2} \,,
                 \label{nuBW} \\
  \lambda_{\rm BW} &= -\frac{{\cal M}^2}{4\pi^2 b^2\rho} \times \nonumber \\
                   & \quad\times
                     \left[(\rho\!+\!b)(E\!-\!K)^2+\frac{(\rho-b)(E\!-\!k'^2 K)^2}{k'^2}\right],
\end{align}
where
\begin{align*}
  &K\equiv K(k) := \int_0^{\pi/2}\frac{{\rm d}\alpha}{\sqrt{1-k^2\sin^2\alpha}} \;, \\
  &E\equiv E(k) := \int_0^{\pi/2}{\sqrt{1-k^2\sin^2\alpha}}\;\,{\rm d}\alpha
\end{align*}
are complete elliptic integrals of the 1st and the 2nd kind,
with modulus and complementary modulus
\[k^2:=1-\frac{(l_1)^2}{(l_2)^2}=\frac{4b\rho}{(l_2)^2}\;, \qquad
  k'^2:=1-k^2=\frac{(l_1)^2}{(l_2)^2} \;.\]
Especially on the axis $\rho=0$, one has $k=0$, $K=E=\pi/2$, so
$\nu_{\rm BW}=-\frac{{\cal M}}{\sqrt{z^2+b^2}}\,$ and $\lambda_{\rm BW}=0$
(the latter must actually hold for {\em any} Weyl solution should the axis be regular).
The solution was derived by \cite{BachW-22} and more recently studied e.g. by \cite{Hoenselaers-95,SemerakZZ-99,DAfonsecaLO-05}.\footnote
{We thank our colleague Pavel \v{C}\'{\i}\v{z}ek for pointing out that we did not give $\lambda_{\rm BW}$ properly in \cite{SemerakZZ-99} and for suggesting a correct form.}

Due to linearity of the Laplace equation, the partial potentials $\nu_{\rm Schw}$ and $\nu_{\rm BW}$ can simply be added, while the total $\lambda$ function has to be found from the total $\nu$ by quadrature.
In Schwarzschild coordinates, the total metric reads \cite{SemerakZZ-99}
\begin{align}
  {\rm d}s^2
  =& -e^{2\nu}{\rm d}t^2+r(r-2M)\,e^{-2\nu}\sin^2\theta\,{\rm d}\phi^2+ \nonumber \\
   & +\left[(r\!-\!M)^2\!-\!M^2\cos^2\theta\right]
      e^{2\lambda-2\nu}\!
      \left[\frac{{\rm d}r^2}{r(r\!-\!2M)}+{\rm d}\theta^2\right] \nonumber \\
  =& -\!\left(1-\frac{2M}{r}\right)e^{2\nu_{\rm ext}}{\rm d}t^2
     +\frac{e^{2\lambda_{\rm ext}-2\nu_{\rm ext}}}{1-\frac{2M}{r}}\,{\rm d}r^2+ \nonumber \\
   & +r^2 e^{-2\nu_{\rm ext}}\!
      \left(e^{2\lambda_{\rm ext}}{\rm d}\theta^2+\sin^2\theta\,{\rm d}\phi^2\right),
  \label{metric}
\end{align}
where in our case $\nu_{\rm ext}(\equiv\nu_1)\equiv\nu_{\rm BW}$,
while $\lambda_{\rm ext}:=\lambda-\lambda_{\rm Schw}=\lambda_{\rm BW}+\lambda_{\rm int}$.
Regarding that
\[\nu_{\rm BW}=-\frac{2{\cal M}}{\pi M}\,\frac{K(k)}{l_2/M} \;,
  \qquad
  \frac{\partial\nu_{\rm BW}}{\partial\rho}=\frac{1}{M}\,\frac{\partial\nu_{\rm BW}}{\partial(\rho/M)}\]
(and similarly for derivatives with respect to $z$ and $r$), we can now add to the decomposition (\ref{lambda-decomp}), on the basis of equations (\ref{lambda_int,eqns}), that $\lambda_{\rm int}$ scales with $M$ and ${\cal M}$ as
\begin{align}
  &\lambda_{\rm int}\!
   \left(\frac{\rho}{M},\frac{z}{M};\,\frac{b}{M};\,M,{\cal M}\right)= \nonumber \\
  &=\frac{{\cal M}}{M}\;\lambda_{\rm int}
    \!\left(\frac{\rho}{M},\frac{z}{M};\,\frac{b}{M};\,M=1,{\cal M}=1\right).
\end{align}
Thanks to this property, one can find the $\lambda$-field for a given system (given $M$, ${\cal M}$, $b$) by simple scaling of its form obtained for $M=1$ and ${\cal M}=1$ (and the given $b$).

\subsection{Behaviour on the horizon}

Our main interest is to learn how the external source affects the geometry inside the black hole, which requires to extend the metric below the horizon. It will thus be useful to know how the metric functions behave on the horizon. In the Weyl coordinates, the horizon is given by $\rho=0$, $|z|\leq M$. The black-hole potential has there a logarithmic divergence while the exterior potential is regular,\footnote
{Asterisk / index `H' denote expansions/values valid at the horizon.}
\[\nu_{\rm Schw}\eqast\ln\frac{\rho}{2\,\sqrt{M^2-z^2}}+O(\rho^2),
  \quad
  \nu_{\rm BW}\eqH -\frac{{\cal M}}{\sqrt{z^2+b^2}} \;,\]
so the total potential $\nu=\nu_{\rm Schw}+\nu_{\rm BW}$ expands there as
\begin{equation}  \label{nu,expand}
  \nu\eqast\ln\frac{\rho}{2\,\sqrt{M^2-z^2}}-\frac{{\cal M}}{\sqrt{z^2+b^2}}+O(\rho^2) \,,
\end{equation}
which implies, for example,
\begin{align}
  \rho^2 e^{-2\nu} &\eqH 4(M^2-z^2)\exp\left(\frac{2{\cal M}}{\sqrt{z^2+b^2}}\right), \\
  \lambda_{,\rho}-\nu_{,\rho} &= \rho(\nu_{,\rho})^2-\rho(\nu_{,z})^2-\nu_{,\rho}\eqast O(\rho) \,.
    \label{lambda,rho-nu,rho}
\end{align}
On any static (in fact even stationary) horizon, $\lambda(z)\eqH 2\nu(z)-2\nu(z\!=\!M)$ (see e.g. \cite{Will-74}, eq. (24)), therefore, applying this for the total as well as pure-Schwarzschild metric, one finds
\begin{align}
  \lambda-\nu &\eqH \lambda_{\rm Schw}-\nu_{\rm Schw}+\nu_{\rm BW}(z)-2\nu_{\rm BW}(z\!=\!M)\eqH {}
              \nonumber \\
              &\eqH  \ln\frac{2M}{\sqrt{M^2-z^2}}-\frac{{\cal M}}{\sqrt{z^2+b^2}}
                       +\frac{2{\cal M}}{\sqrt{M^2+b^2}} \;.  \label{lambda-nu;H}
\end{align}
Using (\ref{lambda,rho-nu,rho}) and (\ref{lambda-nu;H}),
\begin{align}
  \lambda-\nu &(\lambda-\nu)_{\rm H}+\int_0^\rho(\lambda_{,\rho}-\nu_{,\rho})\,{\rm d}\rho
                      \eqast {} \nonumber \\
              &\eqast \ln\frac{2M}{\sqrt{M^2-z^2}}-\frac{{\cal M}}{\sqrt{z^2+b^2}}
                  +\frac{2{\cal M}}{\sqrt{M^2+b^2}}+O(\rho^2)  \label{lambda-nu,expand}
\end{align}
and, by subtraction of (\ref{nu,expand}) from (\ref{lambda-nu,expand}),
\begin{equation}
  \lambda-2\nu\eqast\ln\frac{2M}{\sqrt{M^2-z^2}}-\frac{\rho}{4M}+O(\rho^2) \,.
\end{equation}

\section{Extension of the metric below horizon}
\label{below-horizon}

Interior of a black hole deformed by an external source is known to remain regular, except for the central singularity which however keeps its point-like character \cite{GerochH-82}. In order to extend the metric explicitely, let us first allow the spheroidal radius $r$ to go below $r=2M$. The Schwarzschild potential $\nu_{\rm Schw}$ involves imaginary part ${\rm i}\pi$ there, because the lapse squared $e^{2\nu}$ is negative below horizon. More seriously, the potential induced by the external source has to be continued there since it is not at all defined at that region originally.

\subsection{External potential inside the black hole}

For $r<2M$, the Weyl radius $\rho=\sqrt{r(r-2M)}\,\sin\theta$ turns pure imaginary, which makes the $l_{1,2}$ distances and the modulus of the $K(k)$ integral complex. However, this need not lead to complex $\nu_{\rm BW}$ since the latter is even in $\rho$, as seen, for example, from the known identity
\[K(k)=\frac{2}{1+k'}\,K\!\left(\frac{1-k'}{1+k'}\right)\]
which in our case ($k'=l_1/l_2$) implies
\begin{equation}  \label{K(k)-formula}
  -\frac{\pi}{2{\cal M}}\,\nu_{\rm BW}
  =\frac{K(k)}{l_2}=\frac{2}{l_2+l_1}\,K\!\left(\frac{l_2-l_1}{l_2+l_1}\right).
\end{equation}
This is symmetrical with respect to the exchange $l_1\leftrightarrow l_2$. But such an exchange is equivalent to the change of the sign of $\rho$, so $\nu_{\rm BW}(\rho)$ is even.

Now, if $\nu_{\rm BW}$ is even in $\rho$, it should remain real when $\rho$ becomes pure imaginary. However, the behaviour of $K(k)$ for complex $k^2$ involves a feature which leaves this conclusion only partially valid. Let $\rho$ be pure imaginary, $\rho=:{\rm i}\varrho$, where $\varrho>0$. From the explicit form of the modulus
\begin{equation}
  k^2=\frac{4b\rho}{(l_2)^2}=\frac{4{\rm i}\,b\varrho}{-\varrho^2+b^2+z^2+2{\rm i}\,b\varrho}
\end{equation}
it is seen that inside black hole there is a surface $\varrho^2=b^2+z^2$ where $k^2$ is pure real, $k^2=2$. But $K(k)$ has a branch cut along the real axis at $1<k^2<\infty$, so it is discontinuous on the above surface. More specifically, when crossing the cut from $\Im(k^2)<0$ to $\Im(k^2)>0$ side (which means from $\varrho^2>b^2+z^2$ to $\varrho^2<b^2+z^2$ side of the surface), the integral jumps from $K(k)$ to $K(k)+2{\rm i}\,K(k')$, hence in our case it jumps from $K(\sqrt{2})\doteq 1.311(1-{\rm i})$ to the complex conjugate $K(\sqrt{2})+2{\rm i}\,K({\rm i})\doteq 1.311(1+{\rm i})$. In addition, the same surface also marks the location where $\Re(l_2)=\Im(l_2)$, with $\Re(l_2)<\Im(l_2)$ on its $\varrho^2>b^2+z^2$ side and $\Re(l_2)>\Im(l_2)$ on its $\varrho^2<b^2+z^2$ side.
Due to these two circumstances, the expression $K(k)/l_2$ changes from pure real to pure imaginary when crossing the surface from $\varrho^2<b^2+z^2$ to $\varrho^2>b^2+z^2$.

A possible solution of this issue is offered by the above formula (\ref{K(k)-formula}). Actually, when writing the potential as
\begin{equation}  \label{nuBW,alt}
  \nu_{\rm BW}
  =-\frac{4{\cal M}}{\pi\,(l_2+l_1)}\,K\!\left(\frac{l_2-l_1}{l_2+l_1}\right)
\end{equation}
rather than in the usual form $\nu_{\rm BW}=-2{\cal M}K(k)/(\pi l_2)$, it is real for both real and imaginary $\rho$, it smoothly crosses the horizon and coincides with the original form in the outer region.

Interior solution -- in particular in the region $\varrho^2>b^2+z^2$ where direct extension of the original exterior potential to imaginary $\rho$ did not bring a real result -- can also be checked by returning to the field equations and by solving them once again for $\rho=:{\rm i}\varrho$. The equations then read
\begin{align}
  &\nu_{,\varrho\varrho}+\frac{\nu_{,\varrho}}{\varrho}-\nu_{,zz}=0 \,, \\
  &\lambda_{,\varrho}=\varrho(\nu_{,\varrho})^2+\varrho(\nu_{,z})^2 \,, \quad
   \lambda_{,z}=2\varrho\,\nu_{,\varrho}\nu_{,z} \,,  \label{E-equations,inside}
\end{align}
so in comparison with (\ref{Einstein-eqs}) there appear sign changes in the first two equations. In particular, the first equation is the wave equation in the ``interior meridional plane" $(\varrho,z)$. Its solution, appropriate for our situation, is given by infinite series involving the Legendre functions $P_{n-1/2}\,$:
\begin{align*}
  \nu_{\rm BW}^{\rm in}=
  &-\frac{{\cal M}}{\sqrt{b^2+\varrho^2}} \,\times \\
  &\times \sum\limits_{n=0}^\infty
          \frac{(-1)^n(2n)!}{2^{2n}(n!)^2}\,
          P_{n-\frac{1}{2}}\!\left(\frac{b^2-\varrho^2}{b^2+\varrho^2}\right)
          \frac{z^{2n}}{(b^2+\varrho^2)^n} \;.
\end{align*}
This sum is really an expansion of (\ref{nuBW,alt}) valid inside the horizon.\footnote
{However, it only converges uniformly within $z^2<b^2+\varrho^2$, elsewhere the convergence is just point-wise and slow.}
In particular, on the horizon (more precisely, on all the axis $\varrho=0\Leftrightarrow\rho=0$) it yields correctly
\begin{align}
  \nu_{\rm BW}^{\rm in}(\varrho=0)
     &=-\frac{{\cal M}}{\sqrt{b^2+z^2}}=\nu_{\rm BW}(\rho=0) \,,\\
  \left.\frac{\partial\nu_{\rm BW}^{\rm in}}{\partial\varrho}\right|_{\varrho=0}
     &= 0 =\left.\frac{\partial\nu_{\rm BW}}{\partial\rho}\right|_{\rho=0} \,,\\
  \left.\frac{\partial^2\nu_{\rm BW}^{\rm in}}{\partial\varrho^2}\right|_{\varrho=0}
     &=\frac{{\cal M}}{2}\,\frac{b^2-2z^2}{(b^2+z^2)^{5/2}}
      =-\left.\frac{\partial^2\nu_{\rm BW}}{\partial\rho^2}\right|_{\rho=0}.
\end{align}
An example of the ring-potential behaviour inside the black hole is given in figure \ref{nuBW-plot,inside}.

\subsection{Function $\lambda$ on the axis and at the horizon}

The last function needed in order to complete the metric (\ref{metric}) is $\lambda_{\rm ext}\equiv\lambda-\lambda_{\rm Schw}$. Its extension below the horizon is given by field equations (\ref{E-equations,inside}) which can be rewritten for $\lambda_{\rm ext}$ as
\begin{align}
  \lambda_{{\rm ext},\varrho}
    &\equiv \lambda_{,\varrho}-\lambda_{{\rm Schw},\varrho}
     = \varrho(\nu_{,\varrho})^2+\varrho(\nu_{,z})^2-\lambda_{{\rm Schw},\varrho}= \nonumber \\
    &= \varrho\big[(\nu_{{\rm BW},\varrho})^2+(\nu_{{\rm BW},z})^2 \nonumber \\
    &\qquad         +2\nu_{{\rm Schw},\varrho}\nu_{{\rm BW},\varrho}
                    +2\nu_{{\rm Schw},z}\nu_{{\rm BW},z}\big],
    \label{lambda_ext,varrho} \\
  \lambda_{{\rm ext},z}
    &\equiv \lambda_{,z}-\lambda_{{\rm Schw},z}
     = 2\varrho\,\nu_{,\varrho}\nu_{,z}-\lambda_{{\rm Schw},z}= {} \nonumber \\
    &= 2\varrho\left(\nu_{{\rm Schw},\varrho}\nu_{{\rm BW},z}+\nu_{{\rm Schw},z}\nu_{{\rm BW},\varrho}
                     +\nu_{{\rm BW},\varrho}\nu_{{\rm BW},z}\right).
    \label{lambda_ext,z}
\end{align}
Transforming to the Schwarzschild-type coordinates,
\[\varrho=\sqrt{r(2M-r)}\,\sin\theta \,, \qquad z=(r-M)\cos\theta \,,\]
while now using
\begin{align}
  X_{,r} &= X_{,\varrho}\varrho_{,r}+X_{,z}z_{,r} \nonumber \\
         &=  X_{,\varrho}\;\frac{M-r}{\sqrt{r(2M-r)}}\,\sin\theta+
             X_{,z}\cos\theta \,, \nonumber \\
  X_{,\theta} &= X_{,\varrho}\varrho_{,\theta}+X_{,z}z_{,\theta} \nonumber \\
              &=  X_{,\varrho}\,\sqrt{r(2M-r)}\,\cos\theta+
                  X_{,z}(M-r)\sin\theta \,, \nonumber \\
  \nu_{{\rm Schw},\varrho} &= \frac{(d_1+d_2)\,[4M^2-(d_2-d_1)^2]}
                                   {8M\varrho\;d_1 d_2} \\
                           &=  \frac{M(r-M)\sin\theta}
                                    {[(r-M)^2-M^2\cos^2\theta]\,\sqrt{r(2M-r)}} \;,
      \label{nu_Schw,varrho} \\
  \nu_{{\rm Schw},z} &= \frac{d_2-d_1}{2\,d_1 d_2}
                     = \frac{M\cos\theta}{(r-M)^2-M^2\cos^2\theta}
      \label{nu_Schw,z}
\end{align}
(these formulas are the same as (\ref{nuSchw,rho})--(\ref{nuSchw,z}) valid outside, only $\rho$ is changed for $\varrho$),
the equations assume the form
\begin{align}
  \lambda_{{\rm ext},r}&=
    \frac{2\nu_{{\rm BW},\varrho}\sin\theta}{\sqrt{r(2M-r)}}\,
    \left[\,r(2M-r)\,\nu_{{\rm BW},z}\cos\theta-M\right]
    \nonumber \\ & {}
    +(M-r)\left[(\nu_{{\rm BW},\varrho})^2+(\nu_{{\rm BW},z})^2\right]\sin^2\theta \;, \\
  \lambda_{{\rm ext},\theta}&=
    2\nu_{{\rm BW},z}\sin\theta \;\times \nonumber \\
    &\quad\times
    \left[(M\!-\!r)\sqrt{r(2M\!-\!r)}\,\nu_{{\rm BW},\varrho}\sin\theta\!-\!M\right]
    \nonumber \\ & {}
    +r(2M-r)\left[(\nu_{{\rm BW},\varrho})^2+(\nu_{{\rm BW},z})^2\right]\sin\theta\cos\theta \;.
\end{align}
Note that in Schwarzschild coordinates all the expressions are ``ready to use", whereas if using Weyl coordinates (below horizon), one has to choose the signs of $d_1$ and $d_2$ properly (``by hand") -- see Appendix \ref{appendix-B}.

The first of these reduces, for $\sin\theta=0$, just to
\begin{equation}
  \left(\lambda_{{\rm ext},r}\right)_{\sin\theta=0}=0 \,,
\end{equation}
hence the $\lambda_{\rm ext}$ function is constant along the $\sin\theta=0$ axis. Regarding that on the Weyl axis ($\rho=0$, $|z|>M$) one has $\lambda=\lambda_{\rm Schw}=\lambda_{\rm ext}=0$ ($z={\rm const}$ surfaces are required to be regular there), one thus finds that
\begin{equation}
  \left(\lambda_{\rm ext}\right)_{\sin\theta=0}
  =\left(\lambda_{\rm Schw}\right)_{\sin\theta=0}
  =0
\end{equation}
holds {\em everywhere} on the (Schwarzschild) axis, including the black-hole interior.

Notice now that the second equation for $\lambda_{\rm ext}$ reduces to the same relation at the singularity $r=0$ and on the horizon $r=2M$,
\begin{align}
  \left(\lambda_{{\rm ext},\theta}\right)_{r=0}
     &= -2M\left(\nu_{{\rm BW},z}\right)_{r=0}\sin\theta \,, \\
  \left(\lambda_{{\rm ext},\theta}\right)_{r=2M}
     &= -2M\left(\nu_{{\rm BW},z}\right)_{r=2M}\sin\theta \,.
\end{align}
But $\varrho(r\!=\!2M)=0=\varrho(r\!=\!0)$, $z(r\!=\!2M)=M\cos\theta=-z(r\!=\!0)$ and $\nu_{\rm BW}$ is even in $z$ (hence $\nu_{{\rm BW},z}$ is odd in $z$), so we have
\[\left(\nu_{\rm BW}\right)_{r=0}=\left(\nu_{\rm BW}\right)_{r=2M}, \quad
  \left(\nu_{{\rm BW},z}\right)_{r=0}=-\left(\nu_{{\rm BW},z}\right)_{r=2M}\]
and, therefore,
\begin{equation}
  \left(\lambda_{{\rm ext},\theta}\right)_{r=0}=-\left(\lambda_{{\rm ext},\theta}\right)_{r=2M} \,,
\end{equation}
namely the latitudinal dependence of $\lambda_{\rm ext}$ is just opposite at the singularity and on the horizon.
However, on the horizon we have $\lambda(\theta)\eqH 2\nu(\theta)-2\nu(\theta\!=\!0)$ for the total metric as well as for pure Schwarzschild, so the same must also hold for $\lambda_{\rm ext}\equiv\lambda-\lambda_{\rm Schw}$, hence
\begin{align}
  &\left(\lambda_{\rm ext}\right)_{r=0}=-\left(\lambda_{\rm ext}\right)_{r=2M}= {}  \nonumber \\
  &= -2\nu_{\rm BW}(r\!=\!2M)+2\nu_{\rm BW}(r\!=\!2M,\theta\!=\!0)= {} \nonumber \\ {}
  &= \frac{2{\cal M}}{\sqrt{M^2\cos^2\theta+b^2}}-\frac{2{\cal M}}{\sqrt{M^2+b^2}} \;.
\end{align}
Note that the ``duality" between the horizon and the singularity was already observed by \cite{FrolovS-07}.

\subsection{Function $\lambda$ inside the black hole}

It has thus been possible to find $\lambda$ along the $\sin\theta=0$ axis and on the horizon. One would however like to know its behaviour everywhere inside the black hole. For such a purpose, it has proved advantageous to subtract equations (\ref{lambda_ext,varrho}), (\ref{lambda_ext,z}) and rewrite the result
\[\lambda_{{\rm ext},\varrho}\mp\lambda_{{\rm ext},z}
  =\varrho\,(\nu_{,\varrho}\mp\nu_{,z})^2
   -\varrho\,(\nu_{{\rm Schw},\varrho}\mp\nu_{{\rm Schw},z})^2\]
in terms of the derivatives
\[\frac{\partial}{\partial\eta_\mp}
  :=\frac{\partial}{\partial\varrho}\mp\frac{\partial}{\partial z} \;:\]
\begin{align}
  \lambda_{{\rm ext},\eta_\mp}
  &=\varrho\left[(\nu_{,\eta_\mp})^2-(\nu_{{\rm Schw},\eta_\mp})^2\right]= \nonumber \\
  &=\varrho\left[(\nu_{{\rm Schw},\eta_\mp}+\nu_{{\rm ext},\eta_\mp})^2
                 -(\nu_{{\rm Schw},\eta_\mp})^2\right]= \nonumber \\
  &=\varrho\,\nu_{{\rm ext},\eta_\mp}
    \left(2\nu_{{\rm Schw},\eta_\mp}+\nu_{{\rm ext},\eta_\mp}\right),  \label{lambda_ext,eta}
\end{align}
where, from (\ref{nu_Schw,varrho}) and (\ref{nu_Schw,z}),
\begin{align}
  &\nu_{{\rm Schw},\eta_\mp}
     = 2M\;
       \frac{d_1\left[\varrho\mp(z+M)\right]+d_2\left[\varrho\mp(z-M)\right]}
            {d_1 d_2\left[(d_1+d_2)^2-4M^2\right]} = \nonumber \\
  &\quad = -\frac{M}{2r(2M-r)}\left[\frac{\varrho\mp(z+M)}{d_2}+\frac{\varrho\mp(z-M)}{d_1}\right].
\end{align}
Regarding that
\begin{align*}
  d_{1,2}&=\sqrt{(z\mp M)^2+\rho^2}
          =\sqrt{(z\mp M)^2-\varrho^2}= \\
         &=\sqrt{(z\mp M+\varrho)(z\mp M-\varrho)} \;,
\end{align*}
the above can also be written
\begin{align*}
  &\nu_{{\rm Schw},\eta_\mp}= \\
  &=\pm\frac{M}{2r(2M-r)}
       \left(\sqrt{\frac{z+M\mp\varrho}{z+M\pm\varrho}}
             +\sqrt{\frac{z-M\mp\varrho}{z-M\pm\varrho}}\right).
\end{align*}

%%%%%%%%%%%%%%%%%%%%%%%%%%%%%%%%%%%%%%%%%%%%%%%%%%%%%%%%%%%%%%%%%
\begin{figure}
\centering
\includegraphics[width=\columnwidth]{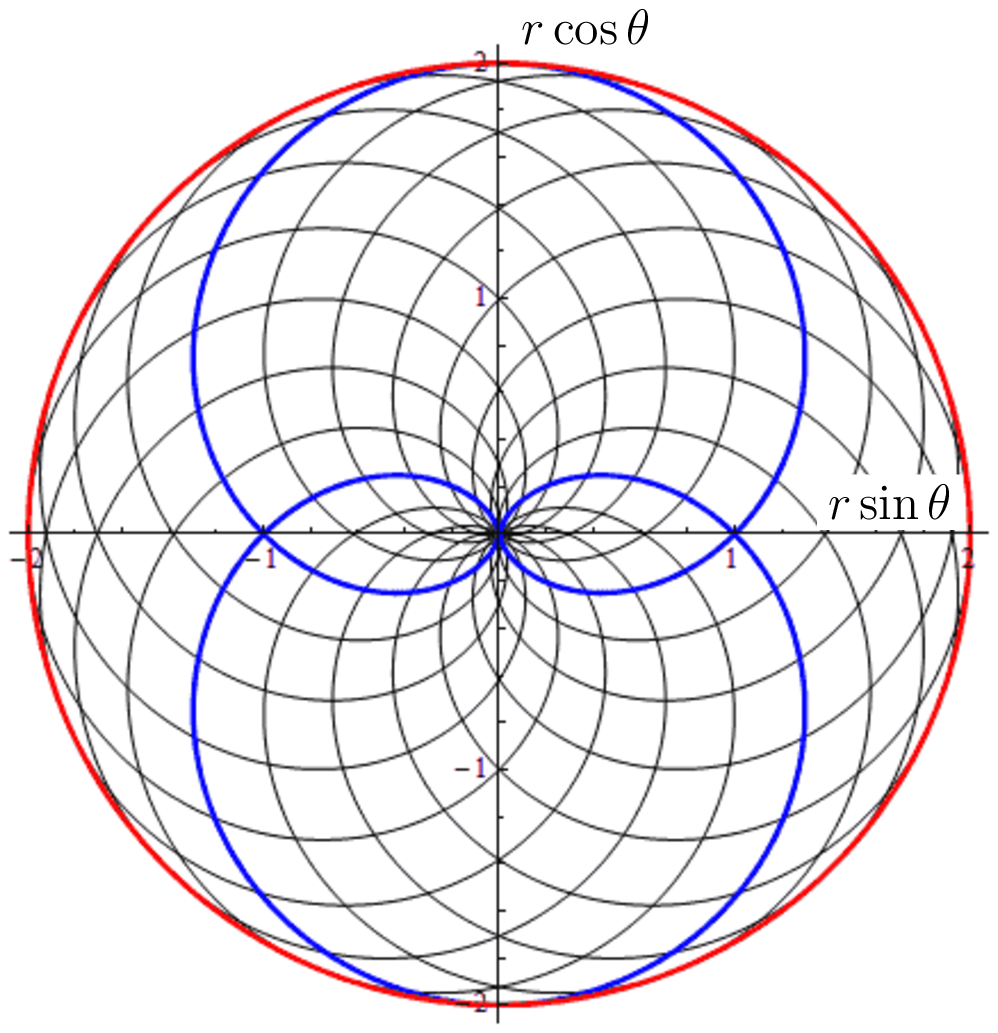}
\caption
{An elegant pattern of null geodesics just tangent to the horizon (red) and spanning the black-hole interior, in the Schwarzschild coordinates given by $r\!=\!M\left[1\pm\cos(\theta\!-\!\theta_0)\right]$ and depicted in $r\sin\theta$, $r\cos\theta$ axes (scaled by $M$); the geodesics given by $\theta_0=0$ and $\theta_0=\pi$ are emphasized (blue). We proceed along these characteristics when integrating the Einstein equations inside the horizon.}
\label{null-geodesics-inside}
\end{figure}
%%%%%%%%%%%%%%%%%%%%%%%%%%%%%%%%%%%%%%%%%%%%%%%%%%%%%%%%%%%%%%%%%

Equation (\ref{lambda_ext,eta}) has now to be integrated toward the black-hole interior.
This is best performed along the family of curves given by
\[d_{1,2}=0 \quad \Leftrightarrow \quad \varrho=|z\mp M|
            \quad \Leftrightarrow \quad r=M(1\pm\cos\theta) \,,\]
namely
\begin{align}
  &r=M\left[1\pm\cos(\theta-\theta_0)\right], \quad
   \theta\in\langle \theta_0,\theta_0+\pi\rangle \,, \label{null-geodesics} \\
  &{\rm where} \quad
   \theta_0={\rm const}\in\langle 0,\pi\rangle \,. \nonumber
\end{align}
These curves are null geodesics starting tangentially to the horizon and descending toward the central singularity (see figure \ref{null-geodesics-inside} and appendix \ref{appendix-A}); they represent characteristics of the Einstein equations.
Multiplying equations (\ref{lambda_ext,eta}) by the tangent vector ${\rm d}\eta_\mp/{\rm d}\sigma$ of the respective curves, where $\sigma$ is some parameter, one obtains an ordinary differential equation suitable for integration,
\begin{equation}
  \frac{{\rm d}\lambda_{\rm ext}}{{\rm d}\sigma}
  = \varrho\,\nu_{{\rm ext},\eta_\mp}\,
    \frac{{\rm d}(2\nu_{\rm Schw}+\nu_{\rm ext})}{{\rm d}\sigma}
    \;.
\end{equation}
The main benefit of the latter is that it no more contains $\nu_{{\rm Schw},\eta_\mp}$ (which does not behave nicely below the horizon).

However, the formulation we have found the most advantageous still requires one more transformation.

\subsection{Horizon angles fixed by characteristics and a trapezoid rule}

It is seen on figure (\ref{null-geodesics-inside}) that any two null geodesics which ``counter-inspiral" (with respect to each other) from the horizon to the singularity intersect at a certain point $(r,\theta)$ inside the black hole. Let us denote by $\theta_+$ and $\theta_-$ the angles on the horizon from where these geodesics start, assuming $0<\theta_-<\theta_+<\pi$, and make the transformation
\begin{align*}
  &r=M\left(1+\cos\frac{\theta_+-\theta_-}{2}\right),
   \quad
   \theta=\frac{\theta_++\theta_-}{2} \;, \\
  &\varrho=\frac{M}{2}\,(\cos\theta_--\cos\theta_+) \,,
   \quad
   z=\frac{M}{2}\,(\cos\theta_-+\cos\theta_+) \,.
\end{align*}
In terms of these angles, the metric reads (notice that it is no longer diagonal)
\begin{align}
  {\rm d}s^2=&-\left(1-\frac{2M}{r}\right)e^{2\nu_{\rm ext}}{\rm d}t^2  \nonumber \\
             &+r^2 e^{-2\nu_{\rm ext}}
               (e^{2\lambda_{\rm ext}}{\rm d}\theta_-{\rm d}\theta_++\sin^2\theta\,{\rm d}\phi^2)\,,
  \label{metric,theta12}
\end{align}
where $r=r(\theta_-,\theta_+)$, and Einstein equations have the form
\begin{align}
  & 2(\cos\theta_-\!-\!\cos\theta_+)\,\frac{\partial^2\nu_{\rm ext}}{\partial\theta_-\partial\theta_+}
    =\frac{\partial\nu_{\rm ext}}{\partial\theta_+}\,\sin\theta_-
     \!-\!\frac{\partial\nu_{\rm ext}}{\partial\theta_-}\,\sin\theta_+ \,, \\
  & \frac{\partial\lambda_{\rm ext}}{\partial\theta_-}\,\sin\theta_-
    =\left[2\sin\theta\!-\!(\cos\theta_-\!-\!\cos\theta_+)\,\frac{\partial\nu_{\rm ext}}{\partial\theta_-}\right]
     \frac{\partial\nu_{\rm ext}}{\partial\theta_-} \;, \label{lambda,theta-} \\
  & \frac{\partial\lambda_{\rm ext}}{\partial\theta_+}\,\sin\theta_+
    =\left[2\sin\theta\!+\!(\cos\theta_-\!-\!\cos\theta_+)\,\frac{\partial\nu_{\rm ext}}{\partial\theta_+}\right]
     \frac{\partial\nu_{\rm ext}}{\partial\theta_+} \;. \label{lambda,theta+}
\end{align}

To solve the first equation, it is sufficient to know the axis values $\nu_{\rm ext}(0,z)$,
\begin{equation}  \label{nuext,integral}
  \nu_{\rm ext}(\varrho,z)=
  \frac{1}{\pi}\int\limits_0^\pi \nu_{\rm ext}(0,z+\varrho\cos\alpha)\,{\rm d}\alpha \,.
\end{equation}
This integral can be calculated using a simple trapezoid rule. Actually, for a function having the same odd derivatives with respect to the integration variable at the end points of the integration interval (which is the case of our $\nu_{\rm ext}(0,z+\varrho\cos\alpha)$), the error of this scheme falls exponentially with the number of discretization points (see e.g. \cite{Numerical-Recipes}, chapter 4).

In order to find $\lambda_{\rm ext}$, we have solved, instead of equations (\ref{lambda,theta-}) and (\ref{lambda,theta+}) themselves, their integrability condition
\begin{equation}
  \frac{\partial^2\lambda_{\rm ext}}{\partial\theta_-\partial\theta_+}=
  \frac{M(\nu_{{\rm ext},\theta_+}-\nu_{{\rm ext},\theta_-})}{2\,\sqrt{r(2M-r)}}
  -\nu_{{\rm ext},\theta_+}\nu_{{\rm ext},\theta_-} \,.
\end{equation}
Using a reversible discretization scheme which respects propagation of the boundary conditions along characteristics (like in numerical treatment of the wave equation), one obtains very precise results, mainly thanks to a regular behaviour everywhere inside the black hole (including the shells where $d_{1,2}=0$).

The language of $\theta_-$ and $\theta_+$ angles is also advantageous for the Kretschmann scalar:
in a vacuum, the Riemann tensor has 3 independent components which satisfy
\begin{align}
  &{R^{t\theta_-}}_{t\theta_-}={R^{t\theta_+}}_{t\theta_+}
   ={R^{\phi\theta_-}}_{\phi\theta_-}={R^{\phi\theta_+}}_{\phi\theta_+}= \nonumber \\
  &\qquad =-\frac{1}{2}\,{R^{t\phi}}_{t\phi}
   =-\frac{1}{2}\,{R^{\theta_-\theta_+}}_{\theta_-\theta_+} \;, \nonumber \\
  &{R^{t\theta_-}}_{t\theta_+}=-{R^{\phi\theta_-}}_{\phi\theta_+} \,, \quad
   {R^{t\theta_+}}_{t\theta_-}=-{R^{\phi\theta_+}}_{\phi\theta_-} \,,
\end{align}
and in terms of which the Kretschmann invariant reads just
\begin{equation}
  K=12\,({R^{\theta_-\theta_+}}_{\theta_-\theta_+})^2+
    16\,{R^{t\theta_-}}_{t\theta_+}{R^{t\theta_+}}_{t\theta_-} \;.
\end{equation}

\section{Potential, field and curvature inside and outside black hole}
\label{invariants}

%%%%%%%%%%%%%%%%%%%%%%%%%%%%%%%%%%%%%%%%%%%%%%%%%%%%%%%%%%%%%%%%%
\begin{figure*}
\centering
\includegraphics[width=\textwidth]{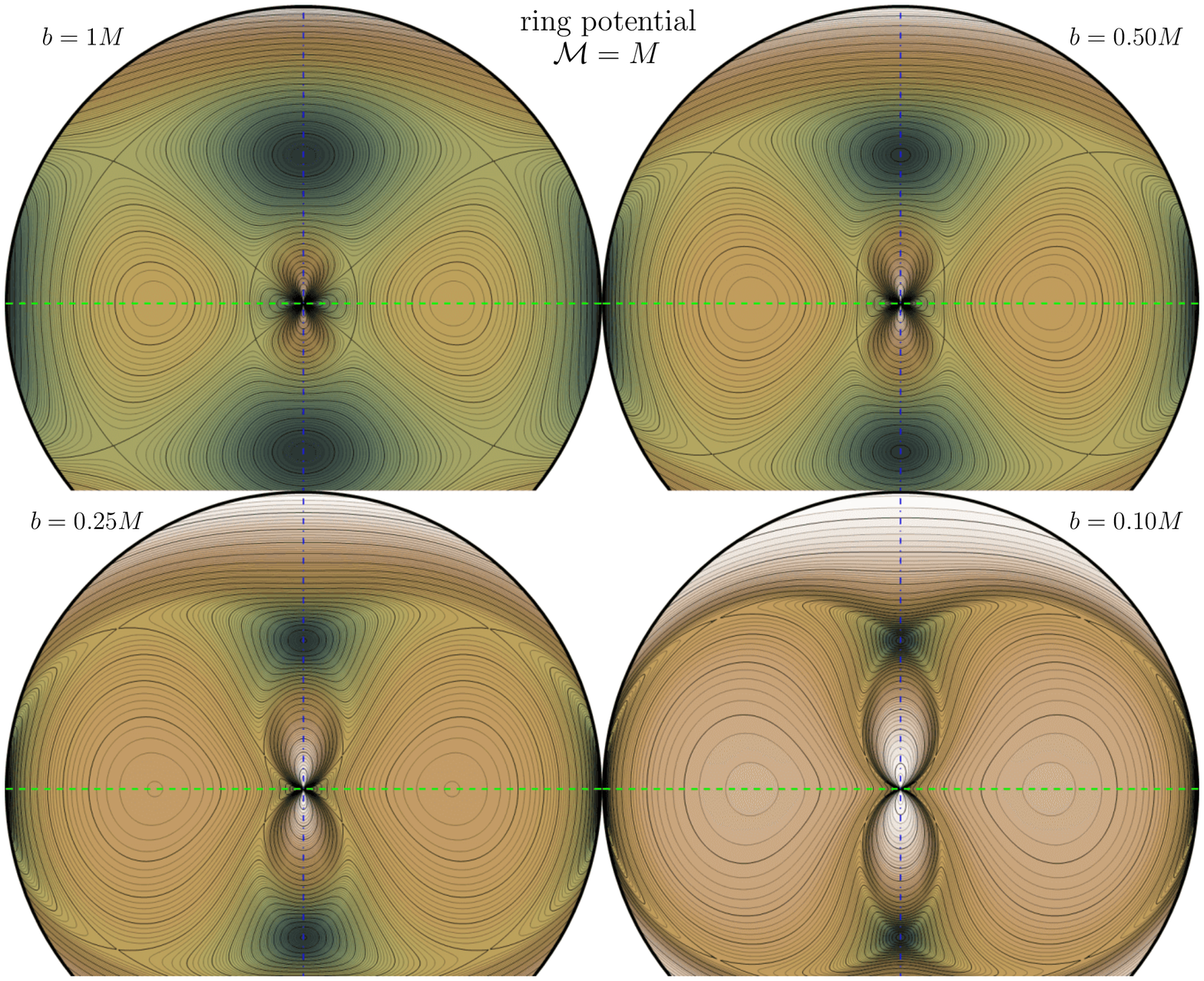}
\caption
{Meridional-plane contours of the BW-ring potential $\nu_{\rm BW}$ inside a black hole, plotted for a ring of mass ${\cal M}=M$ and of different Weyl radii $b$ (given in the plots). The plots are drawn in Schwarzschild-type coordinates, so they are spherical and symmetric with respect to the equatorial plane (where the ring is placed) indicated by the green dashed line, as well as with respect to the axis indicated by the dot-dashed blue line. Higher/lower values correspond to brown/green colour.}
\label{nuBW-plot,inside}
\end{figure*}
%%%%%%%%%%%%%%%%%%%%%%%%%%%%%%%%%%%%%%%%%%%%%%%%%%%%%%%%%%%%%%%%%

%%%%%%%%%%%%%%%%%%%%%%%%%%%%%%%%%%%%%%%%%%%%%%%%%%%%%%%%%%%%%%%%%
\begin{figure*}
\centering
\includegraphics[width=\textwidth]{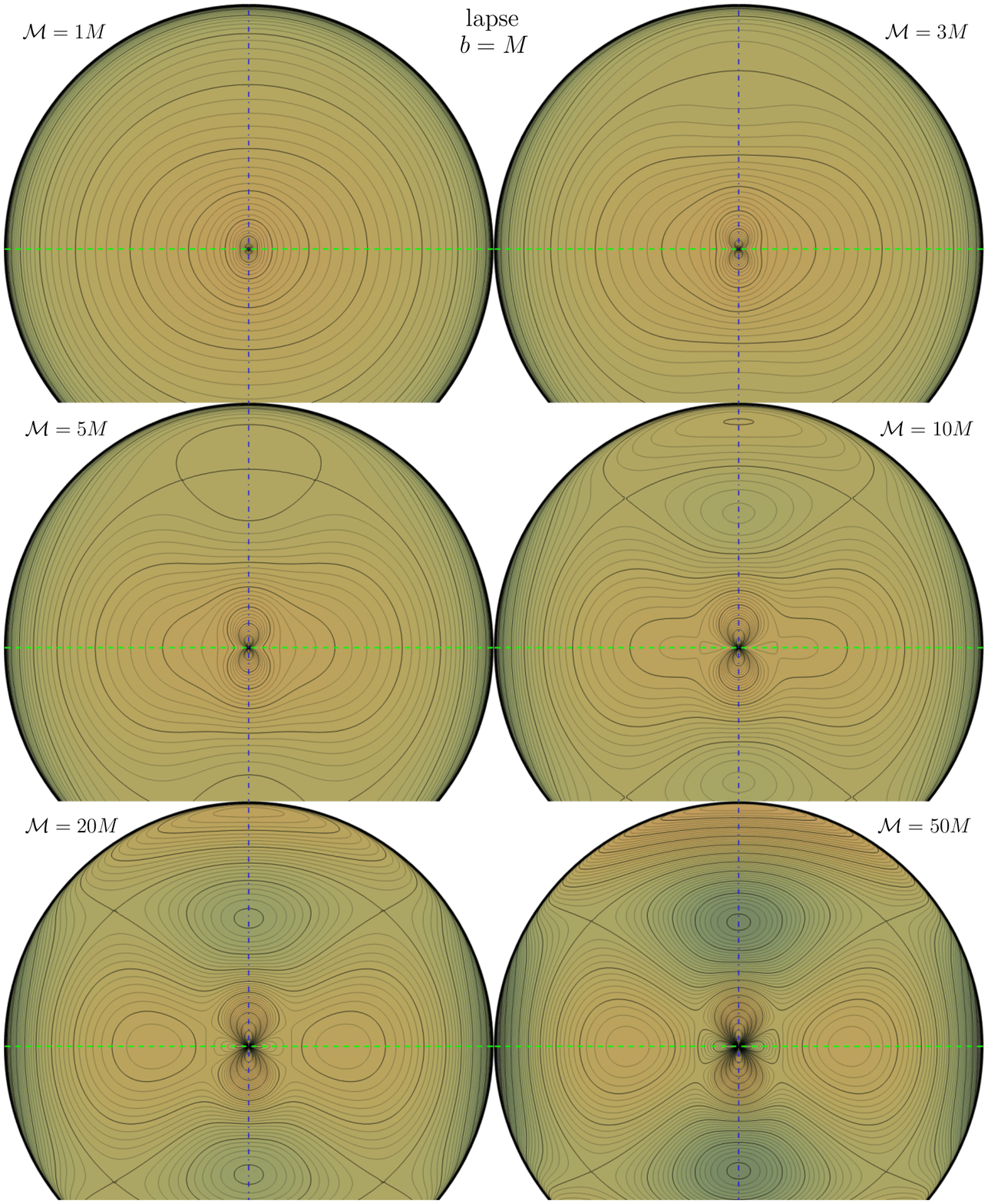}
\caption
{Meridional-plane contours of lapse $N$ (or of potential $\nu$) inside a black hole surrounded by a BW ring with radius $b\!=\!M$ and of different masses ${\cal M}$ (given in the plots). The plots are drawn in Schwarzschild-type coordinates, so they are spherical and symmetric with respect to the equatorial plane (where the ring is placed) indicated by the green dashed line, as well as with respect to the axis indicated by the dot-dashed blue line. Note that on the horizon $N$ vanishes.}
\label{N-mass}
\end{figure*}
%%%%%%%%%%%%%%%%%%%%%%%%%%%%%%%%%%%%%%%%%%%%%%%%%%%%%%%%%%%%%%%%%

%%%%%%%%%%%%%%%%%%%%%%%%%%%%%%%%%%%%%%%%%%%%%%%%%%%%%%%%%%%%%%%%%
\begin{figure*}
\centering
\includegraphics[width=\textwidth]{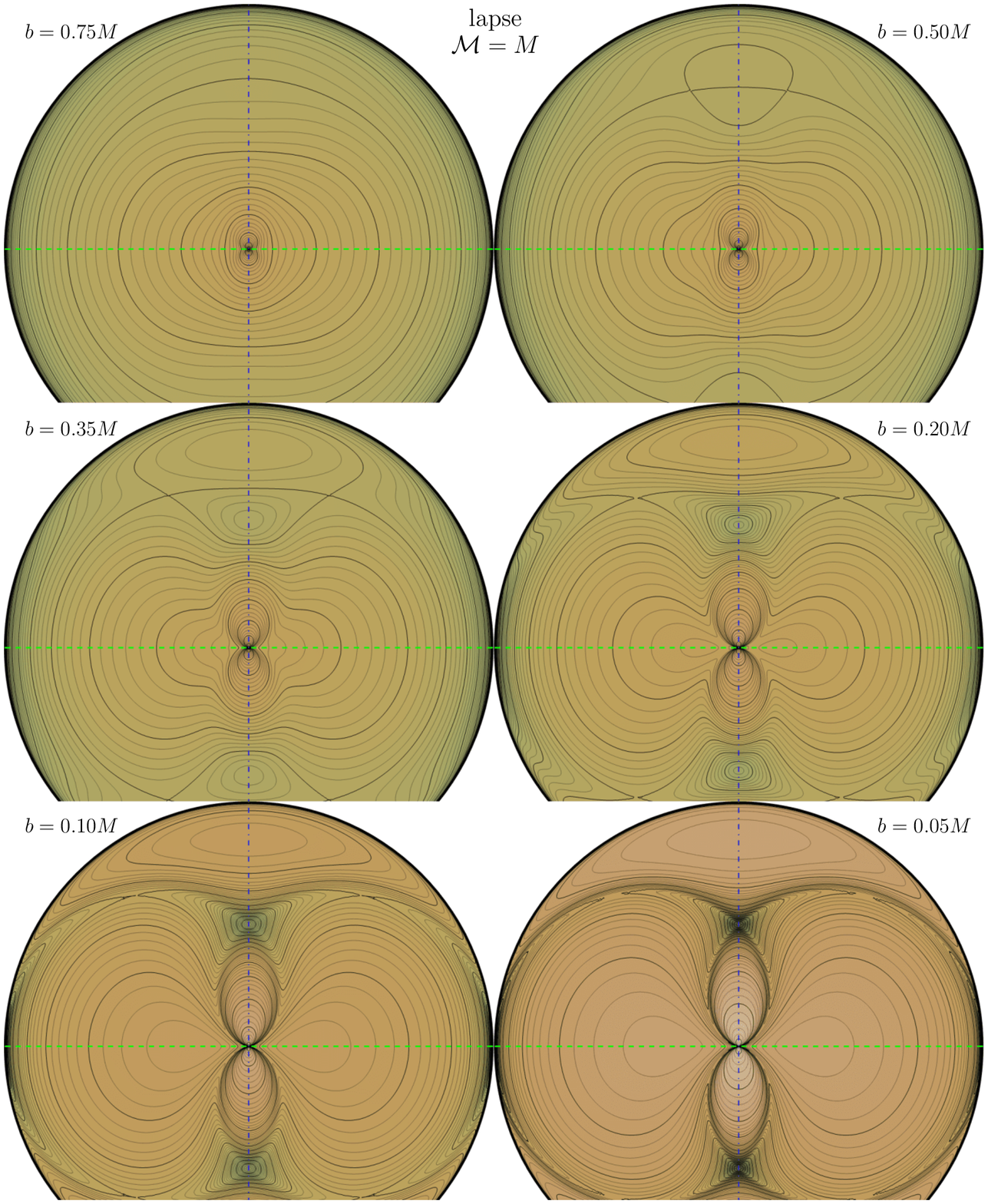}
\caption
{Meridional-plane contours of lapse $N$ (or of potential $\nu$) inside a black hole surrounded by a BW ring of mass ${\cal M}\!=\!M$ and of different Weyl radii $b$ (given in the plots). Meaning of the plots is the same as in figure \ref{N-mass}. Inside the black hole (originally spherically symmetric), local minima (more green) and maxima (more brown) clearly develop due to the surrounding ring. In the axial region they are of spheroidal shape, while in the equatorial region they are toroidal.}
\label{N-radius}
\end{figure*}
%%%%%%%%%%%%%%%%%%%%%%%%%%%%%%%%%%%%%%%%%%%%%%%%%%%%%%%%%%%%%%%%%

%%%%%%%%%%%%%%%%%%%%%%%%%%%%%%%%%%%%%%%%%%%%%%%%%%%%%%%%%%%%%%%%%
\begin{figure*}
\centering
\includegraphics[width=\textwidth]{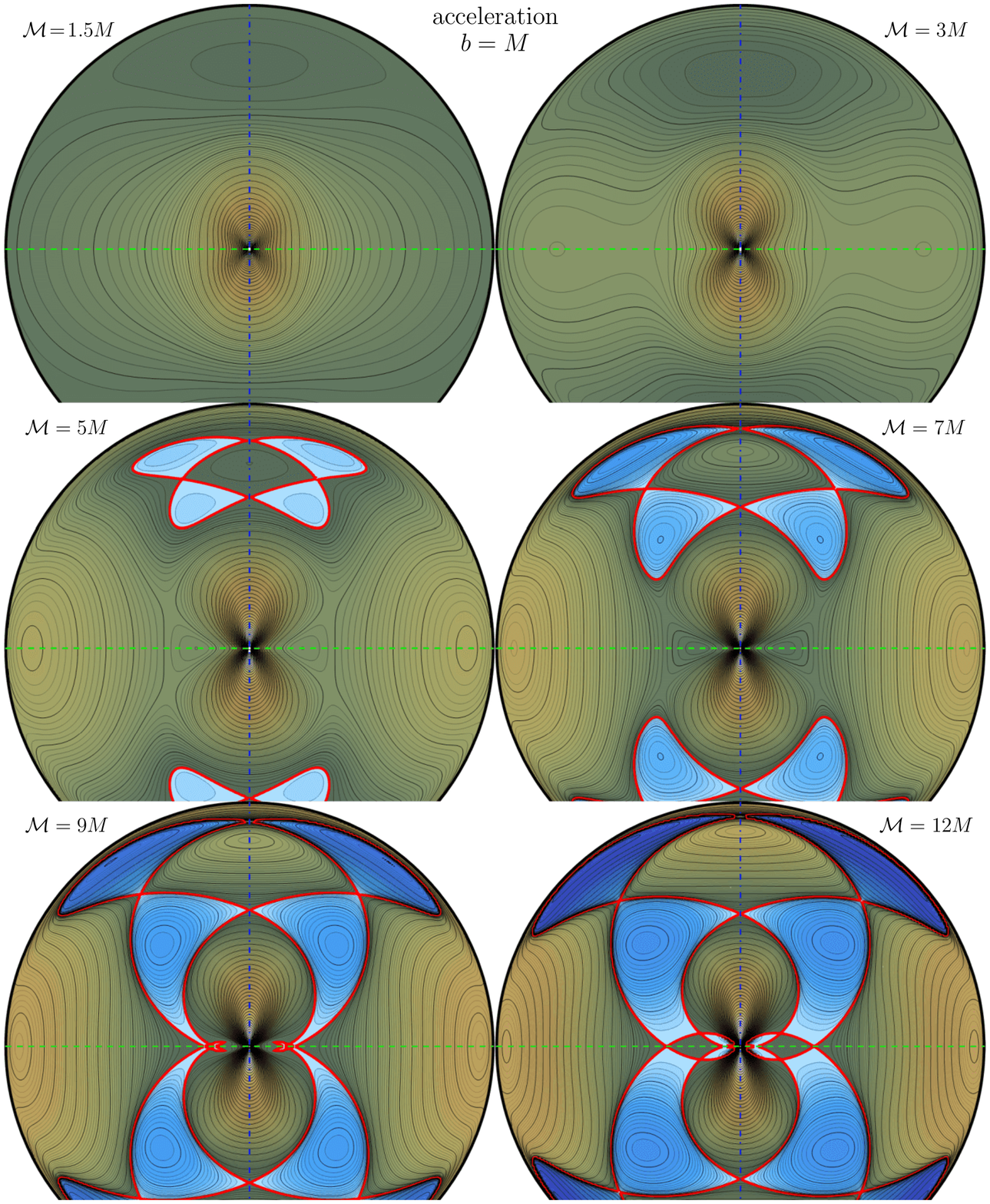}
\caption
{Meridional-plane contours of gravitational acceleration (squared) $\kappa^2$ inside a black hole surrounded by a BW ring with radius $b\!=\!M$ and of different masses ${\cal M}$ (given in the plots). Meaning of the plots is the same as in previous figures showing lapse. Again brown/green indicates higher/lower value. Drawn in blue with red boundaries are the regions of {\em negative} values, where ``acceleration of a static observer" is time-like. On the horizon $\kappa$ assumes a uniform value.}
\label{kappa-mass}
\end{figure*}
%%%%%%%%%%%%%%%%%%%%%%%%%%%%%%%%%%%%%%%%%%%%%%%%%%%%%%%%%%%%%%%%%

%%%%%%%%%%%%%%%%%%%%%%%%%%%%%%%%%%%%%%%%%%%%%%%%%%%%%%%%%%%%%%%%%
\begin{figure*}
\centering
\includegraphics[width=\textwidth]{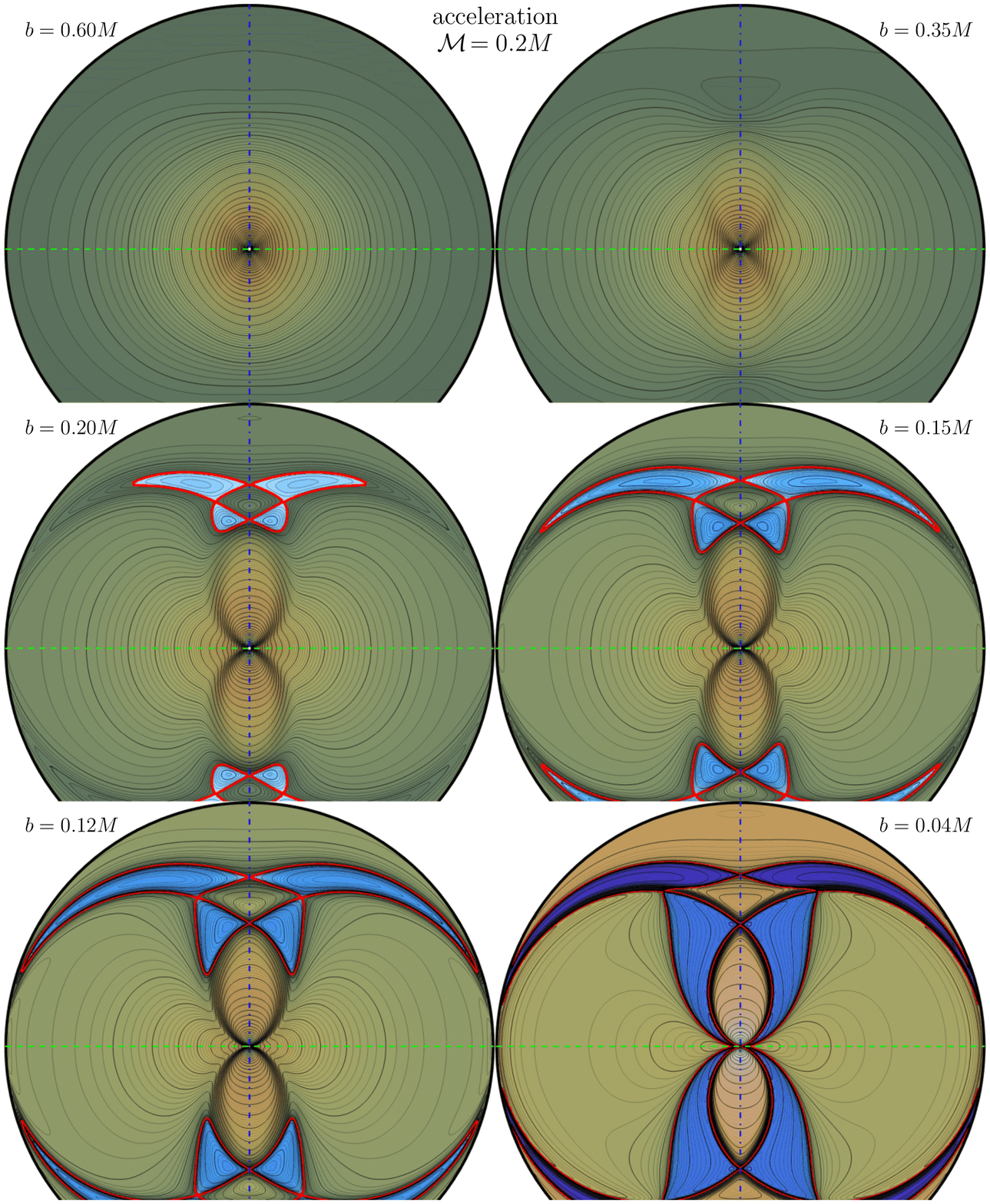}
\caption
{Meridional-plane contours of gravitational acceleration (squared) $\kappa^2$ inside a black hole surrounded by a BW ring of mass ${\cal M}\!=\!0.2M$ and of different Weyl radii $b$ (given in the plots). Meaning of the plots is the same as in previous figures. The blue regions of time-like ``acceleration" again develop into quite a complicated arrangement (the quotation marks just remind that below horizon this quantity does not have its usual sense).}
\label{kappa-radius}
\end{figure*}
%%%%%%%%%%%%%%%%%%%%%%%%%%%%%%%%%%%%%%%%%%%%%%%%%%%%%%%%%%%%%%%%%

Similarly as in the first paper on Majumdar--Papapetrou black-hole binary, we reveal the space-time geometry on the behaviour of the simplest invariants given by the metric and its first and second derivatives. Here, however, we deal with {\em vacuum} solution, so the Ricci tensor is zero and it makes no sense to study its quadratic scalar. We will thus consider the lapse function $N=e^\nu$, the gravitational acceleration $\kappa$ given by $\kappa^2=g^{\mu\nu}N_{,\mu}N_{,\nu}$ and the Kretschmann scalar $K=R_{\mu\nu\kappa\lambda}R^{\mu\nu\kappa\lambda}$.

Since all the configurations are static, axially symmetric and reflectionally symmetric with respect to the ``equatorial plane" (the one in which the ring is placed), we show their properties on meridional plots with Schwarzschild coordinates $(r\sin\theta,r\cos\theta)$ (in which the horizon is a sphere on $r=2M$). In all the figures, a ``geographic" colouring is used, with brown/green indicating higher/lower positive values and light/dark blue indicating smaller/greater depths.
In figure \ref{nuBW-plot,inside}, the sole ring potential $\nu_{\rm BW}$ inside the black hole is shown first, for ${\cal M}=M$ and several different ring radii. The potential has nothing special at the horizon and is also regular everywhere below it. However, the figure shows that inside the black hole it propagates\footnote
{This verb reminds that the black-hole interior is a {\em dynamical} domain.}
in a non-trivial manner.
The lapse-function $N\!=\!\sqrt{(2M/r)-1}\;e^{\nu_{\rm BW}}$ contours inside the horizon are shown, for sequences of black-hole--ring space-times, in figures \ref{N-mass} (fixed ring radius $b=M$, increasing mass) and \ref{N-radius} (fixed ring mass ${\cal M}=M$, decreasing radius). Their shapes clearly follow the behaviour of the external, ring potential. The colouring is not so ``attractive" as in the following figures of acceleration and curvature, simply because the values of $N$ are not so extreme; they only fall to zero (dark green) on the horizon (very suddenly) and, interestingly, also at certain locations on the axis.

The gravitational-acceleration level contours are shown in figures \ref{kappa-mass} and \ref{kappa-radius}. The pattern is rather different from that of potential/lapse, involving quite a complicated arrangement of regions where $\kappa^2$ is negative (drawn in blue). This means that the gradient of lapse, which {\em outside} the horizon determines acceleration of static observers (those at rest with respect to infinity)\footnote
{Let us also remind that at the horizon $\kappa$ is known as {\it surface gravity} and that over stationary horizons it is constant.}
and is everywhere space-like there, becomes time-like in some interior zones if the ring is sufficiently ``strong".

%%%%%%%%%%%%%%%%%%%%%%%%%%%%%%%%%%%%%%%%%%%%%%%%%%%%%%%%%%%%%%%%%
\begin{figure*}
\centering
\includegraphics[width=\textwidth]{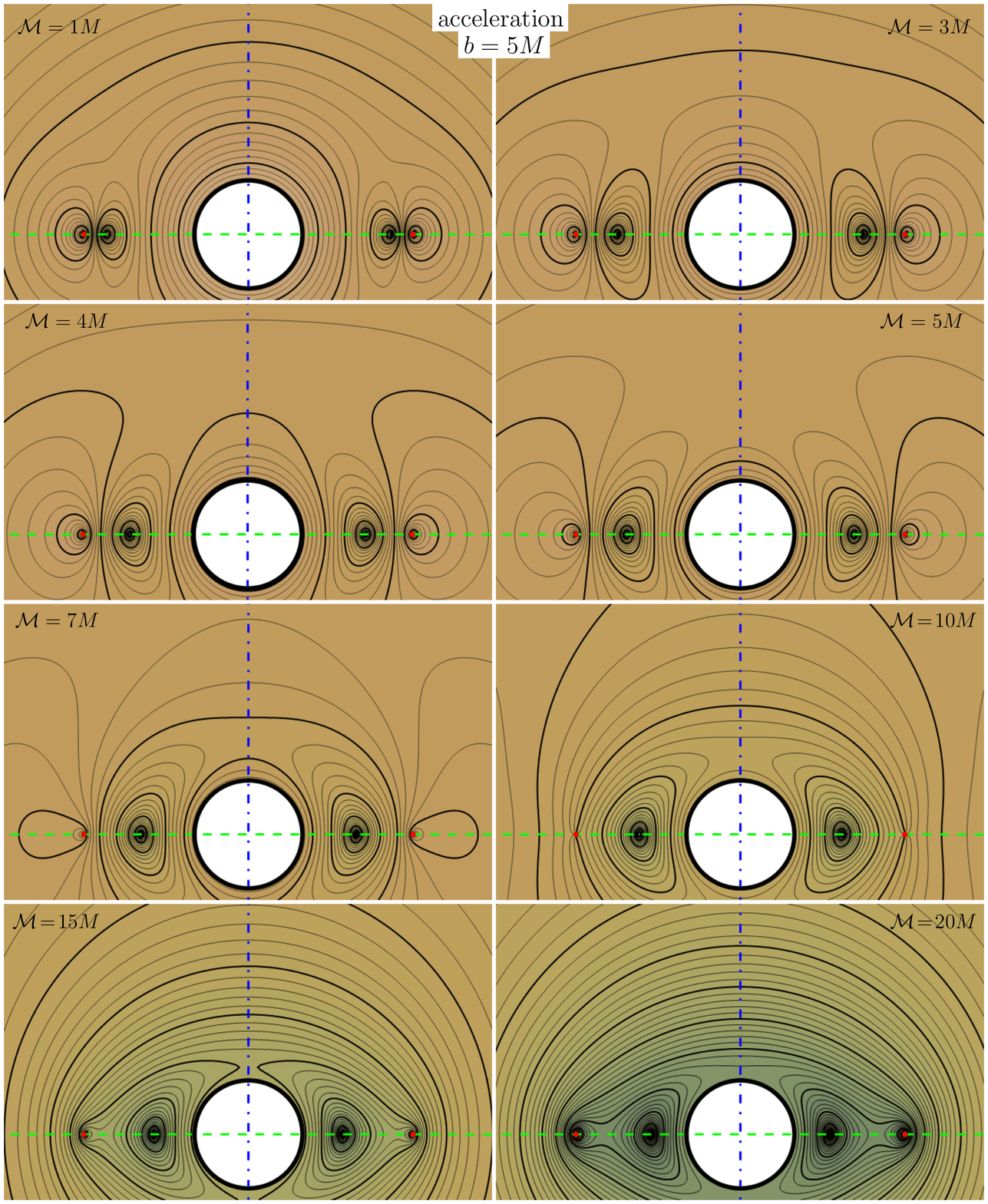}
\caption
{Meridional-plane contours of $\kappa^2$ outside a black hole surrounded by a BW ring (red dots) with radius $b\!=\!5M$ (Schwarzschild radius $\doteq 6.1M$) and of different masses ${\cal M}$ (given in the plots). The white circular region (of radius $r\!=\!2M$) is the black hole, which indicates scale of the plots. The most distinct feature is the quite sharp zero (unstable-equilibrium ring) between the ring and the hole, shifting towards the horizon with gradual increase of the ring mass. On the horizon $\kappa\!=\!{\rm const}$.}
\label{kappa-mass-out}
\end{figure*}
%%%%%%%%%%%%%%%%%%%%%%%%%%%%%%%%%%%%%%%%%%%%%%%%%%%%%%%%%%%%%%%%%

We have not taken much notice of the geometry {\em outside} of the horizon, mainly focusing on deformation of the interior. At the level of potential ($\nu$, or the lapse function $N$), it would be rather superfluous to present figures of the exterior, because these simply correspond to the Newtonian potential of a finite rod, surrounded, symmetrically, by a thin ring and transformed from the Weyl to the Schwarzschild coordinates. The level of field (acceleration) may already be more interesting, since, admittedly, that is not a common transformation from cylindrical to spherical coordinates. In figure \ref{kappa-mass-out}, we thus show how the acceleration ($\kappa^2$) field changes with mass of the ring when the latter is placed on $b=5M$ (which corresponds to $r\doteq 6.1M$ in terms of the Schwarzschild radius). No surprise is seen, in particular, no intriguing structure along the axis; the main feature is the ring of unstable equilibrium (zero acceleration) between the ring and the horizon, gradually shifting from the former to the latter while the ring mass is being increased.

\subsection{The Kretschmann scalar}
\label{Kretschmann}

%%%%%%%%%%%%%%%%%%%%%%%%%%%%%%%%%%%%%%%%%%%%%%%%%%%%%%%%%%%%%%%%%
\begin{figure*}
\centering
\includegraphics[width=\textwidth]{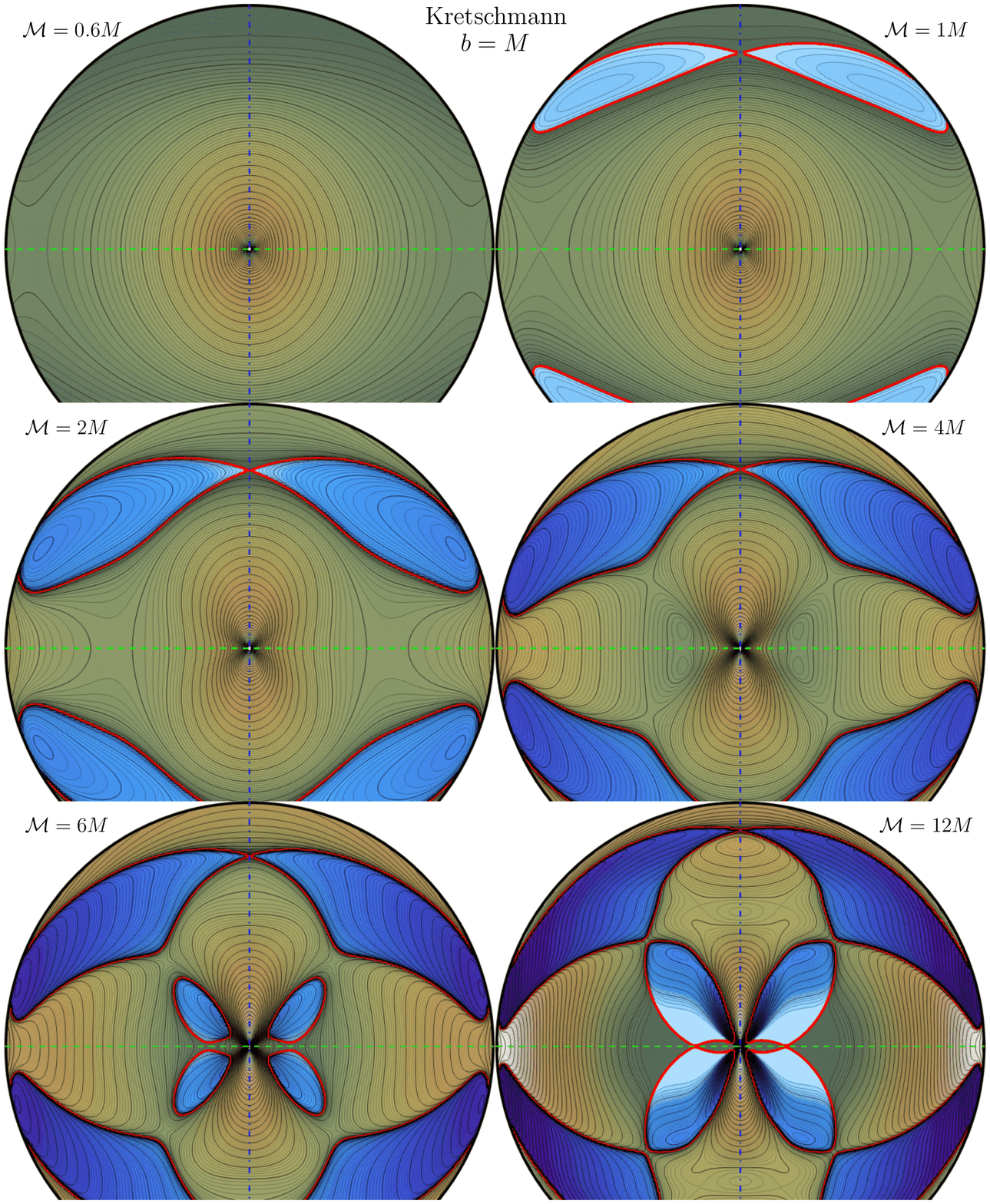}
\caption
{Meridional-plane contours of the Kretschmann scalar inside a black hole surrounded by a BW ring with radius $b\!=\!M$ and of different masses ${\cal M}$ (given in the plots). Meaning of the plots is the same as in previous figures. The scalar is {\em negative} in the regions drawn in blue with red boundary; it is seen that these regions need not always touch the horizon.}
\label{Kretschmann-mass}
\end{figure*}
%%%%%%%%%%%%%%%%%%%%%%%%%%%%%%%%%%%%%%%%%%%%%%%%%%%%%%%%%%%%%%%%%

%%%%%%%%%%%%%%%%%%%%%%%%%%%%%%%%%%%%%%%%%%%%%%%%%%%%%%%%%%%%%%%%%
\begin{figure*}
\centering
\includegraphics[width=\textwidth]{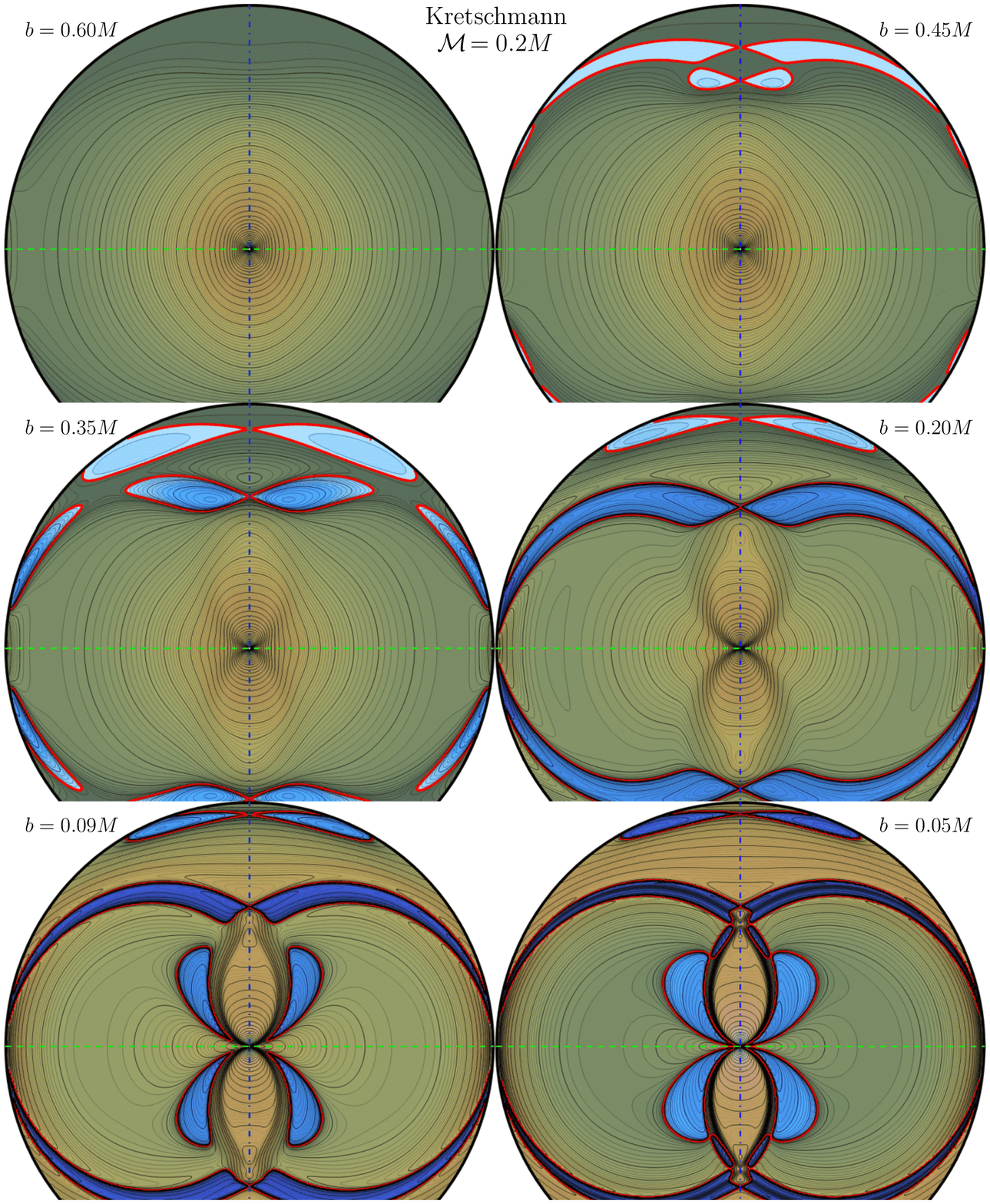}
\caption
{Meridional-plane contours of the Kretschmann scalar inside a black hole surrounded by a BW ring with mass ${\cal M}\!=\!0.2M$ and of different Weyl radii $b$ (given in the plots). Meaning of the plots is the same as in previous figures. The negative-value regions (blue with red border) form a complicated structure, mainly around the central singularity.}
\label{Kretschmann-radius}
\end{figure*}
%%%%%%%%%%%%%%%%%%%%%%%%%%%%%%%%%%%%%%%%%%%%%%%%%%%%%%%%%%%%%%%%%

Finally we turn to the Kretschmann invariant. In the preceding paper \cite{SemerakB-16} (equation (33)), we used the Weyl-coordinate expression to compute it. This time, specifically in the dynamical region inside the black hole, the Schwarzschild-coordinate form is more suitable, for computation as well as for interpretation. It is quite similar,
\begin{align}
  \!\!\!K
  &= 8\left[({R^{tr}}_{tr})^2\!+\!({R^{t\theta}}_{t\theta})^2\!+\!({R^{t\phi}}_{t\phi})^2
            \!-\frac{2\,({R^{tr}}_{t\theta})^2}{r(2M-r)}\right]  \label{K-Schw} \\
  &= \frac{8e^{4\nu-4\lambda}}{r^6} \;\times {} \nonumber \\
  &  \quad\times
     \left[(\tilde{R}^{tr}{}_{tr})^2\!+\!(\tilde{R}^{t\theta}{}_{t\theta})^2
           \!+\!(\tilde{R}^{t\phi}{}_{t\phi})^2
           \!-\frac{2r\,(\tilde{R}^{tr}{}_{t\theta})^2}{2M-r}\right], \nonumber
\end{align}
where we have denoted ($j=r,\theta,\phi$, no summation)
\[\tilde{R}^{tj}{}_{tj}:=r^3 e^{2\lambda-2\nu}{{R}^{tj}}_{tj} \;,
  \quad
  \tilde{R}^{tr}{}_{t\theta}:=r^2 e^{2\lambda-2\nu}{{R}^{tr}}_{t\theta} \;.\]
Several simple observations:
\begin{itemize}
\item
For pure Schwarzschild, one has $e^{4\nu-4\lambda}=1$, $\tilde{R}^{tr}{}_{tr}=2M$, $\tilde{R}^{t\theta}{}_{t\theta}=M$, $\tilde{R}^{t\phi}{}_{t\phi}=M$ and $\tilde{R}^{tr}{}_{t\theta}=0$, which yields $K=48M^2/r^6$ correctly.
\item
$K$ is fully determined by the ``electric-type" tidal field, as expected in a static space-time. The first three components are related by vacuum Einstein equations, ${{R}^{tk}}_{tk}=R^t_t=0$ (summation over $k$).
\item
$K>0$ everywhere outside the black hole ($r>2M$). Inside, it can only become negative due to the $R^{tr}{}_{t\theta}$ component, i.e. the off-diagonal component of the ``electric" tidal field (which vanishes in pure Schwarzschild).
\item
As the external-source potentials are regular at $r=0$, the strong singularity of this central point is not altered by them.
\end{itemize}

%%%%%%%%%%%%%%%%%%%%%%%%%%%%%%%%%%%%%%%%%%%%%%%%%%%%%%%%%%%%%%%%%
\begin{figure*}
\centering
\includegraphics[width=\textwidth]{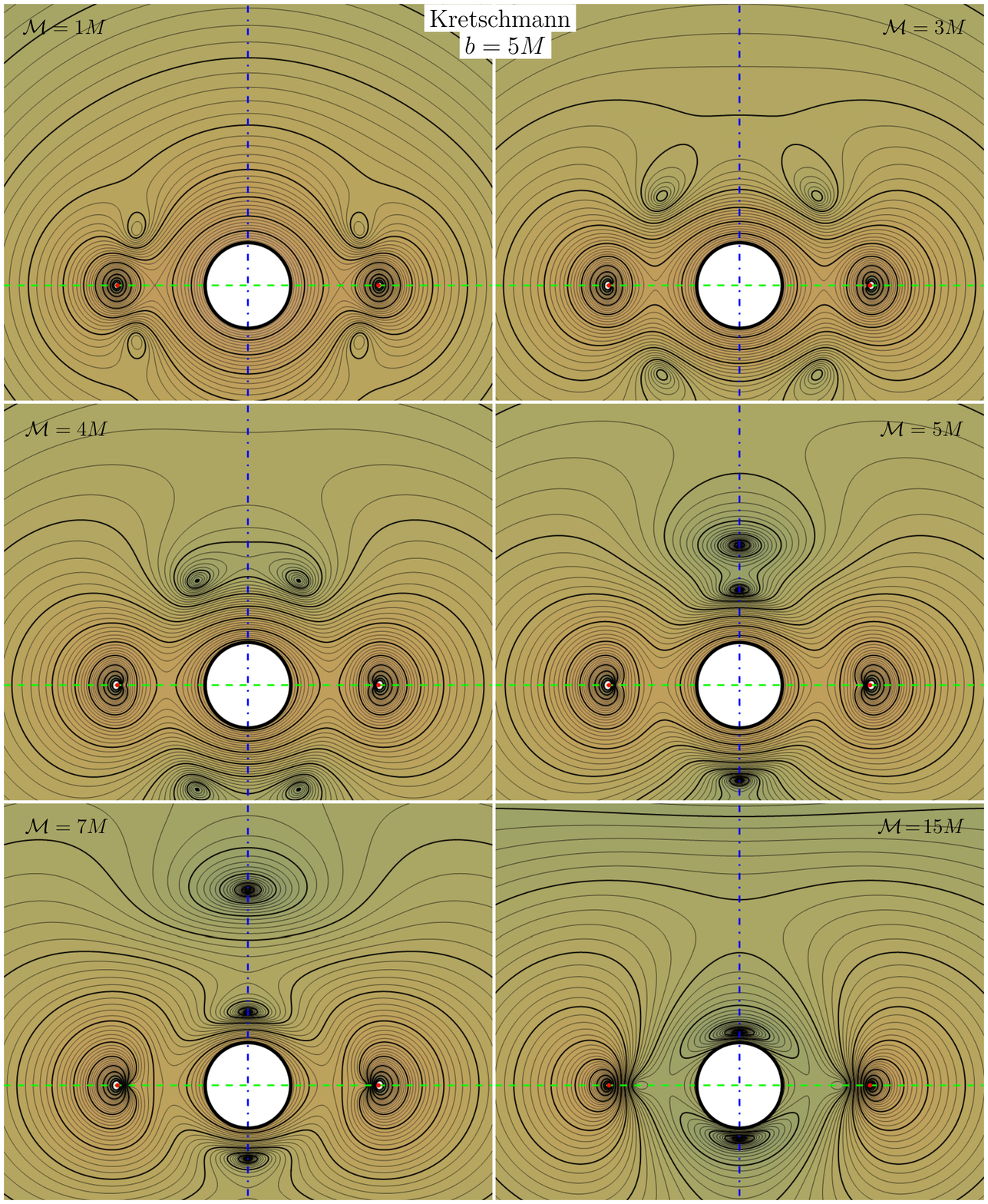}
\caption
{Meridional-plane contours of the Kretschmann scalar outside a black hole surrounded by a BW ring (red dots at divergent maximum of the invariant) with radius $b\!=\!5M$ ($r\doteq 6.1M$) and of different masses ${\cal M}$ (given in the plots). The white circular region (of radius $r\!=\!2M$) is the black hole, which indicates scale of the plots. Notice mainly the ring-shaped minima rising from the BW ring, approaching the axis and then splitting into one receding and one approaching the horizon.}
\label{Kretschmann-mass-out}
\end{figure*}
%%%%%%%%%%%%%%%%%%%%%%%%%%%%%%%%%%%%%%%%%%%%%%%%%%%%%%%%%%%%%%%%%

As opposed to the lapse and gravitational acceleration, the Kretschmann scalar requires knowing ``the second" metric function $\lambda$ which has to be found numerically by integrating Einstein's equations. Outside the black hole, one standardly follows some vacuum line starting from the axis (where $\lambda=0$). The results are illustrated in figure \ref{Kretschmann-mass-out} (ring at $b=5M$, or $r\doteq 6.1M$, sequence showing dependence on the ring mass). Besides the saddle ring, expected between the Bach-Weyl ring and the horizon (very slowly shifting towards the horizon with increasing ring mass), quite an interesting feature can be seen off the equatorial plane: with the ring mass increasing from zero, a ring-shaped minimum raises from the ring and goes ``up" (and also down, symmetrically, of course) while deepening and shrinking in radius; for a certain ring mass, it shrinks to the very axis, and then splits into two profound minima, of which one continues to recede along the axis, while the second approaches the horizon ``northern pole".

However, we have mainly focused on the black-hole interior again. In order to determine $\lambda$ by integration of the field equations, we have followed there the characteristics given by null geodesics starting tangentially to the horizon, as described at the end of section \ref{below-horizon}. The results are shown in figures \ref{Kretschmann-mass} (dependence on the ring mass) and \ref{Kretschmann-radius} (dependence on the ring's Weyl radius). Both sequences show that the curvature inside horizon is influenced considerably. Typically, with increasing strength of perturbation due to the ring, the regions of negative Kretschmann scalar occur and develop in a non-trivial manner; we draw them in blue and indicate their borders by red lines. Let us stress that the masses and radii chosen are out of any astrophysically realistic range, in order to mainly see how the interior curvature behaves under extremely strong perturbations. (The latter may only apply to a system of a black hole surrounded by a very compact, neutron torus, which might occur -- very temporarily -- during the collapse of a compact binary.)

%%%%%%%%%%%%%%%%%%%%%%%%%%%%%%%%%%%%%%%%%%%%%%%%%%%%%%%%%%%%%%%%%
\begin{figure*}
\centering
\includegraphics[width=0.75\textwidth]{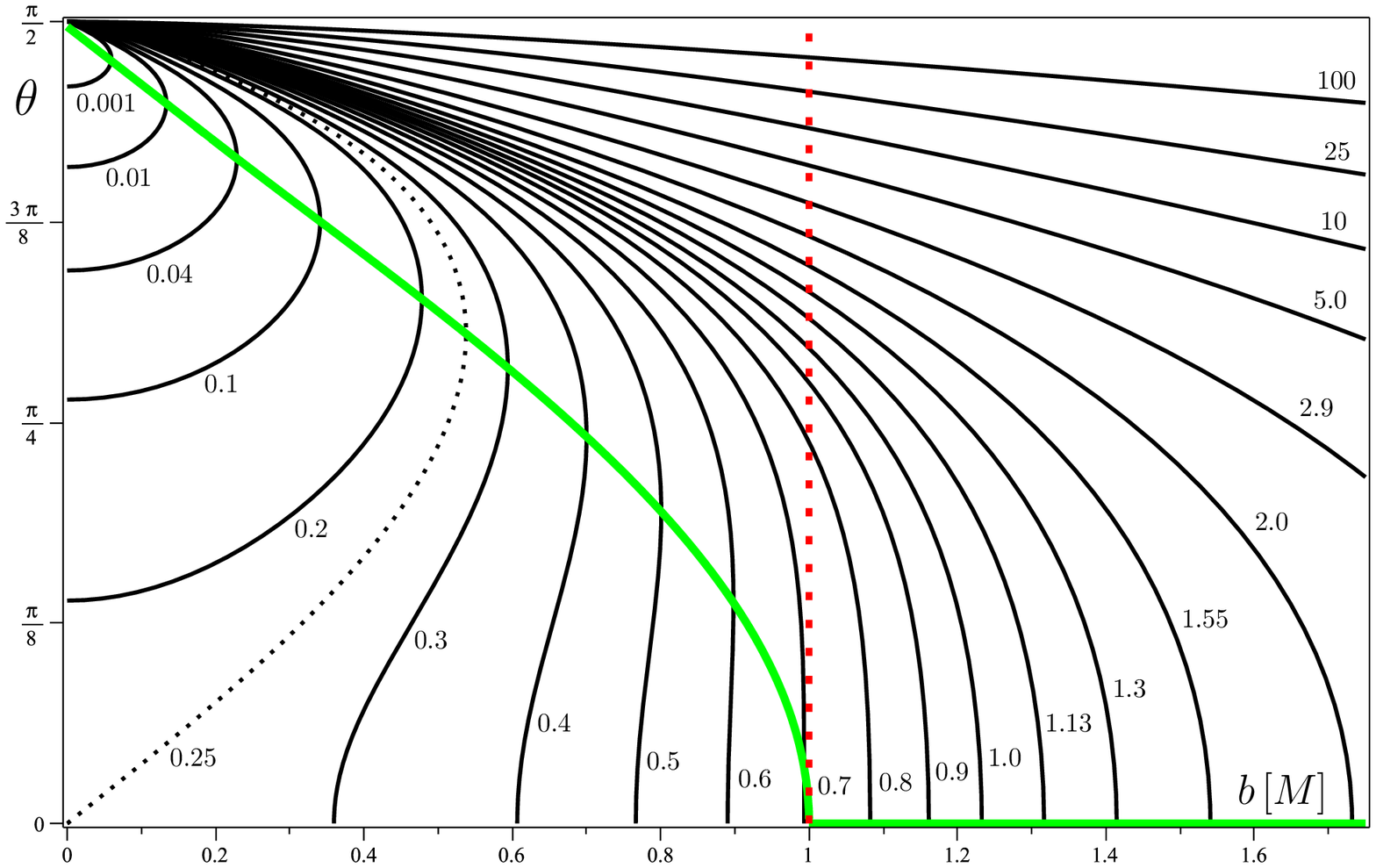}
\caption
{Curves showing zeros of the horizon's Gauss curvature within the $(b,\theta)$ plane, in dependence on mass of the Bach-Weyl ring ${\cal M}$. As this mass is increased from zero, the zero curve expands from just tiny loop at $b\rightarrow 0$, $\theta\rightarrow\pi/2$ toward bottom and right. Specifically, the curves shown correspond to ${\cal M}/M=0.001$, 0.01, 0.04, 0.1, 0.2, 0.3, 0.4, 0.5, 0.6, 0.7, 0.8, 0.9, 1.0, 1.13, 1.30, 1.55, 2.0, 2.9, 5.0, 10, 25, 100, as indicated in the plot.
The figure is to be used as follows: choose the ring's Weyl radius $b$ and mass ${\cal M}$; then, the Gauss curvature of the horizon is negative at latitudes $\theta$ where the given $b={\rm const}$ line lies to the left of the (black) curve corresponding to the chosen ${\cal M}$; where it lies to the right of the latter, the Gauss curvature is positive. It is seen that for any given $b$, there always exist a certain minimal value of ${\cal M}$ from which there appears a region of negative Gauss curvature on the horizon. If $b\geq M$, such a region starts spreading from the axis ($\theta=0$), whereas if $b<M$, it opens out from some {\em non-axial} latitude; this latitude shifts toward the equatorial plane with $b$ decreasing from $M$ to zero. The locations where, for a given $b$, the Gauss curvature vanishes first (in increasing the mass) and then gets negative, are marked by solid green line. It is also seen -- one verifies the exact value from equation (\ref{Gauss,axis}) -- that for ${\cal M}$ below $M/4$ (dotted black), the Gauss curvature can never get negative at the axis (though it does become negative somewhere closer to the equatorial plane if $b$ is sufficiently small, namely less than $0.5377M$). Equation (\ref{Gauss,axis}) also implies that the important point $b=M$, $\theta=0$ is reached by the curve obtained for ${\cal M}=M/\sqrt{2}$.}
\label{Gauss=0}
\end{figure*}
%%%%%%%%%%%%%%%%%%%%%%%%%%%%%%%%%%%%%%%%%%%%%%%%%%%%%%%%%%%%%%%%%

In order to more understand zeros of the Kretschmann scalar, we recall the relation \cite{FrolovS-07}
\begin{equation}
  K\eqH 3\,{^{(2)}\!}R^2
\end{equation}
between the (four-dimensional) Kretschmann invariant and the Gauss curvature ${^{(2)}\!}R/2$ of the horizon (we mean of the horizon's $t={\rm const}$ section; ${^{(2)}\!}R$ denotes the corresponding 2D Ricci scalar). The Gauss curvature reads, for a generic static axisymmetric source (``ext"), (see e.g. Erratum of \cite{SemerakZZ-99})
\begin{align}
  \frac{{^{(2)}\!}R}{2}
  &\eqH -\,\frac{1}{\sqrt{g_{\theta\theta}g_{\phi\phi}}}
           \left[\frac{(\sqrt{g_{\phi\phi}})_{,\theta}}
                      {\sqrt{g_{\theta\theta}}}\right]_{,\theta} \nonumber \\
  &\eqH -\,\frac{1}{R_{\rm H}^{2}e^{\lambda_{\rm ext}}\sin\theta}
           \left[\frac{(R_{\rm H}\sin\theta)_{,\theta}}
                      {R_{\rm H}e^{\lambda_{\rm ext}}}\right]_{,\theta} \nonumber \\
  &\eqH \frac{1+(\nu_{{\rm ext},\theta}+\lambda_{{\rm ext},\theta})\cot\theta
                 +\nu_{{\rm ext},\theta\theta}-\nu_{{\rm ext},\theta}\lambda_{{\rm ext},\theta}}
               {R_{\rm H}^{2}e^{2\lambda_{\rm ext}}} \nonumber \\
  &\eqH \frac{1+3\nu_{{\rm ext},\theta}\cot\theta
                 +\nu_{{\rm ext},\theta\theta}-2(\nu_{{\rm ext},\theta})^2}
               {4M^{2}\exp\left(2\nu_{\rm ext}(\theta)\!-\!4\nu_{\rm ext}(0)\right)} \;,
\end{align}
where $r=2M$ everywhere and
\begin{equation}
  R_{\rm H}:=2Me^{-\nu_{\rm ext}(r=2M,\theta)}
\end{equation}
is the horizon's equatorial circumferential radius
($2\pi R_{\rm H}\sin\theta$ is its proper azimuthal circumference at given $\theta$ and
$2\int_{0}^{\pi}\!R_{\rm H}(\theta)\,e^{\lambda_{\rm ext}(r=2M,\theta)}{\rm d}\theta$
is its proper poloidal circumference).
For the Bach-Weyl ring, the horizon's Gauss curvature is always positive in the equatorial plane, namely
\begin{equation}
  {^{(2)}\!}R\,(\theta\!=\!\pi/2) \eqH
  \frac{1+\frac{{\cal M}M^2}{b^3}}
       {2M^2\exp\left(\frac{4{\cal M}}{\sqrt{b^2+M^2}}-\frac{2{\cal M}}{b}\right)} \;,
\end{equation}
whereas on the axis it may assume both signs,
\begin{equation}  \label{Gauss,axis}
  {^{(2)}\!}R\,(\theta\!=\!0) \eqH
  \frac{1-\frac{4{\cal M}M^2}{(b^2+M^2)^{3/2}}}
       {2M^2\exp\left(\frac{2{\cal M}}{\sqrt{b^2+M^2}}\right)} \;.
\end{equation}

The horizon is known to get more and more oblate when the exterior source lying in the equatorial plane grows in mass. Intuition and experience would suggest that the horizon's Gaussian curvature mainly tends to zero and then to negative values in the axial region (see e.g. Erratum of \cite{SemerakZZ-99} for the Schwarzschild black hole affected by the concentric BW ring or thin annular disc, and \cite{Semerak-02} for a stationary generalization), but figure \ref{Gauss=0} shows that it is only so for $b>M$. When increasing the mass of a ring lying at $b<M$ (very close to the horizon), the region of negative Gauss curvature opens from some {\em non-axial} location; see the green line in figure \ref{Gauss=0} which indicates the latitude where this happens: it actually shifts toward the equatorial plane if the ring is placed closer and closer to the horizon ($b\rightarrow 0$). With figure \ref{Gauss=0} in mind, one understands better the configuration of negative-Kretschmann regions in figures \ref{Kretschmann-mass} and \ref{Kretschmann-radius}, because they touch the horizon exactly where the latter's Gauss curvature vanishes.

Let us also notice ``why" (or at least when/where) the Kretschmann scalar turns negative. Curvature components are probably most straightforwardly interpreted from the geodesic-deviation equation (see e.g. \cite{NiZ-78} for a detailed interpretation of the geodesic-deviation terms in a proper reference frame of a physical observer).
For those present in (\ref{K-Schw}), it is useful to regard the deviation's Schwarzschild components
\begin{align}
  \frac{{\rm D}^2\delta t}{{\rm d}\tau^2}
  =& -\left[{R^t}_{rtr}(u^r)^2\!+
            {R^t}_{\theta t\theta}(u^\theta)^2\!+
            {R^t}_{\phi t\phi}(u^\phi)^2\right] \delta t  \nonumber \\
   & +\left[{R^t}_{rtr}u^r \delta r+
            {R^t}_{\theta t\theta}u^\theta\delta\theta+
            {R^t}_{\phi t\phi}u^\phi\delta\phi\right] u^t  \nonumber \\
   & +{R^t}_{rt\theta}(u^r\delta\theta+u^\theta\delta r)\,u^t \nonumber \\
   & -2{R^t}_{rt\theta}\,u^r u^\theta\,\delta t \,, \\
  \frac{{\rm D}^2\delta\theta}{{\rm d}\tau^2}
  =& -\left[{R^\theta}_{t\theta t}(u^t)^2\!+
            {R^\theta}_{r\theta r}(u^r)^2\!+
            {R^\theta}_{\phi\theta\phi}(u^\phi)^2\right] \delta\theta  \nonumber \\
   & +\left[{R^\theta}_{t\theta t}u^t \delta t+
            {R^\theta}_{r\theta r}u^r\delta r+
            {R^\theta}_{\phi\theta\phi}u^\phi\delta\phi\right] u^\theta  \nonumber \\
   & +{R^\theta}_{trt}(u^r\delta t-u^t\delta r)\,u^t  \nonumber \\
   & +{R^\theta}_{\phi r\phi}(u^r\delta\phi-u^\phi\delta r)\,u^\phi \,.
\end{align}
For easier intuition, consider a couple of particles separated just in radius $t$ (we are below horizon!), so with $\delta x^i=0\,$ at a given point:
\begin{align}
  \frac{{\rm D}^2\delta t}{{\rm d}\tau^2}
  =& -\left[{R^t}_{rtr}(u^r)^2\!+
            {R^t}_{\theta t\theta}(u^\theta)^2\!+
            {R^t}_{\phi t\phi}(u^\phi)^2\right] \delta t  \nonumber \\
   & -2{R^t}_{rt\theta}\,u^r u^\theta\,\delta t \,, \\
  \frac{{\rm D}^2\delta\theta}{{\rm d}\tau^2}
  =& \left({R^\theta}_{trt}u^r+{R^\theta}_{t\theta t}u^\theta\right) u^t\delta t \,.
\end{align}
Similarly, for particles separated only by $\delta\theta$,
\begin{align}
  \frac{{\rm D}^2\delta t}{{\rm d}\tau^2}
  =& \left({R^t}_{rt\theta}u^r+{R^t}_{\theta t\theta}u^\theta\right)u^t\delta\theta \,, \\
  \frac{{\rm D}^2\delta\theta}{{\rm d}\tau^2}
  =& -\left[{R^\theta}_{t\theta t}(u^t)^2\!+
            {R^\theta}_{r\theta r}(u^r)^2\!+
            {R^\theta}_{\phi\theta\phi}(u^\phi)^2\right] \delta\theta \,.
\end{align}
Hence, the diagonal electric-type components (contributing positively to the Kretschmann scalar) are those which have the particles accelerate relative to each other in directions in which these are already separated, thus causing their longitudinal {\it expansion/contraction} (these terms are always non-zero, because, in the brackets, at least $u^r$ must be so). In contrast, the off-diagonal electric-type components (specifically ${R^t}_{rt\theta}\sim {R^\theta}_{trt}$, contributing negatively to the scalar) are seen to be pulling -- for example -- in $\theta$ the particles separated in $t$ direction et vice versa, thus causing transversal {\it shear}.
Well, note the diagonal terms ${R^\theta}_{t\theta t}u^\theta u^t\delta t$ in $\frac{{\rm D}^2\delta t}{{\rm d}\tau^2}$ and its counter-part ${R^t}_{\theta t\theta}u^\theta u^t\delta\theta$ in $\frac{{\rm D}^2\delta t}{{\rm d}\tau^2}$, and, on the other hand, the non-diagonal term $2{R^t}_{rt\theta}\,u^r u^\theta\,\delta t$ in $\frac{{\rm D}^2\delta t}{{\rm d}\tau^2}$: these do not seem to fit in the above division. But these terms require, besides the separation $\delta t$, also some transverse velocity (namely $u^\theta$); without this velocity, they vanish.

\section{Concluding remarks}
\label{concluding}

Continuing the study of black holes deformed by some additional source, we have found that the presence of a thin ring (described by the Bach--Weyl solution) affects the black-hole field much more than the presence of ``the other" black hole within the Majumdar--Papapetrou binary solution (considered in paper I). Outside the horizon, the potential (lapse) and field (acceleration) behave in a rather Newtonian manner, while curvature (the Kretschmann scalar) displays rather  rugged landscape with loops or points of deep minima developing in the sources' vicinity and changing with parameters.

In the black hole interior the situation is yet more complex.
The gravitational acceleration, given by gradient of lapse/potential, shows different shapes and in extreme situations (very strong ring effect) may become time-like (the corresponding scalar $\kappa^2$ may turn negative).
The curvature is influenced by the ring even more profoundly.\footnote
{It is worth noting, however, that in spite of the considerable effect seen on the black hole, it has been shown by \cite{Guerlebeck-15}, on behaviour of multiple moments, that at infinity it still looks like Schwarzschild (it has ``no hair" there induced by tidal deformation).}
If the ring is placed sufficiently close to the horizon (and/or is sufficiently massive), there even appear two or more toroidal regions of negative Kretschmann scalar $K$. Some of them touch the horizon at circles where the 2D-horizon's Gauss curvature changes sign from positive to negative values. If the Riemann tensor is split into electric and magnetic parts (see e.g. \cite{CostaN-13}), the negative values of $K$ are naturally interpreted as regions where magnetic curvature dominates \cite{CherubiniBCR-02}. However, magnetic effects are usually tied to rotation, whereas here no rotation is present in space-time (though we are dealing with a non-extreme black-hole interior now, so {\em not} with a {\em static} region, of course).
It would be interesting to also study the rich curvature structure inside the ring-perturbed black hole by the scalar-gradient method pursued in \cite{AbdelqaderL-12} or the vortex-tendex concepts suggested by \cite{Nichols-etal-11}.

An obvious remark should be added to visualization. The Schwarzschild-type coordinates we have been using are favourable since they represent the horizon spherical irrespectively of the external influence. However, before interpreting the picture obtained in (any) coordinates, one should bear in mind that most of the statements made are coordinate dependent and that the true geometrical relations may be significantly different. In our case, this does not only apply to the shapes of those various equi-surfaces, but also e.g. to the ``location" (radius) of the ring. Actually, the proper radius
\[\sqrt{g_{rr}}\,{\rm d}r=\frac{e^{\lambda_{\rm ext}-\nu_{\rm ext}}}{\sqrt{1-\frac{2M}{r}}}\;{\rm d}r\]
as well as the circumferential radius
\[\sqrt{g_{\phi\phi}}=r\,\sin\theta\,e^{-\nu_{\rm ext}}\]
depend on the external source; specifically, for a given $r$ they both rapidly grow with the source mass. (Consequently, when ``keeping the ring's radius $b$ while increasing its mass" in the figures, one effectively makes the ring larger and larger, thus {\em weakening} its effect on the black hole.)
This aspect of curvature could be overcome by representing the surfaces in terms of isometric embedding to ${\mathbb E}^3$, which in our (axially symmetric) case means by drawing azimuthal circumferential radius as ``$x$"-axis and proper distance in the meridional direction, but, unfortunately, the shapes provided by strong-field geometry are often very weird and not even reasonably embeddable (have negative curvature). The Bach--Weyl ring, after all, is at finite proper distance from outside, but infinitely far when approached from below, its proper circumference being infinite from either side.

Let us mention some options for future work. First, we saw in paper I on Majumdar--Papapetrou binary that an extreme black hole is not in every respect a strong source, and this paper II confirmed that much stronger effect has been created by a singular ring. An interesting curvature structure might also be offered, within the same class of static and axially symmetric space-times, by a black-hole binary supported by an Appell ring. Namely, this ring generates a field which is ``repulsive" in a certain region, which might be enough to held the holes apart (without any struts). Another possibility is a black-hole binary whose components are held from infinity by singular struts. Such a system (a zero-acceleration limit of the C-metric) is of course artificial and, as opposed to the black hole surrounded by a ring, can hardly approximate any astrophysical setting, but (i) its black holes are far (actually, as far as possible) from extreme state, so they can be expected to exert more strain to each other (than the extreme ones), and (ii) the region between the central singularities of the holes may be rather ``unspoilt" by the singularities stretched along the ``exterior" parts of the symmetry axis. One might also consider a similar system made of a black hole and a massive particle(s). Alternatively, one might subject a black hole to a strong (electro-)magnetic field, e.g. within the Ernst class of exact solutions, but such a field is likely to produce much weaker space-time deformation than the above compact sources.

It would certainly be interesting to extend the analysis to {\em stationary} (non-static, rotating) situations. Although practically tractable and physically sound exact superpositions are not yet available within this class, one could describe them by multipole expansions and study the effect of the individual terms then. In the static (originally Schwarzschild) case, the effect of multipoles has recently been considered by \cite{AbdolrahimiMT-15} in order to learn how they deform the shadow of the horizon; the stationary (originally Kerr) case has been treated by \cite{Shoom-15,PaniGMF-15,AbdolrahimiKNT-15}.\footnote
{Even more general is the dynamical situation, recently studied by \cite{OSullivanH-14,OSullivanH-15}, for example.}
(See \cite{Johannsen-16} for the astrophysical importance of such studies, especially connected with the observational challenge provided by the compact object in our Galactic center.)
The tidal deformation of black holes has also been treated perturbatively, following many routes, see for instance \cite{PoissonV-10,Poisson-15} and references therein.

When speaking of static versus stationary settings, we should recall once more that the interior of the above-considered black hole is {\em dynamical},\footnote
{This is in contrast to the Majumdar-Papapetrou--type black holes studied in the first paper, which are extreme and so in their case the Killing vector field $\partial x^\mu/\partial t$ is time-like everywhere except on the very horizon.}
in order to stress again that all the results obtained ``below horizon" factually describe the $t\!=\!{\rm const}$ sections of the interior. Since the conformal diagram of the hole \& ring space-time is like that of the Schwarzschild black hole alone, just with the equatorial version having a singularity along $r\!=\!r_{\rm ring}$, it is clear how these (time-like) sections look in such a diagram, and that the dynamics ``happens" in the direction of decreasing $r$ (from the horizon towards the singularity) on them. Due to the time symmetry (which however is space-like below the horizon), these sections all have the same geometry. 
It may also be a future plan to compare the results with those obtained for a different slicing of the black-hole interior, in particular for a space-like one (e.g. that defined by constant Kruskal-like time coordinate), which would reveal the dynamics of the interior in a different manner.

Let us conclude by noting that recently we have been studying the black-hole--disc/ring system for another but related reason as well: due to perturbation by the additional source, even within such highly symmetric space-times as static and axisymmetric (also reflection symmetric) ones, the geodesic dynamics in the black-hole field looses complete integrability and may incline to chaos (see \cite{SukovaS-13} and preceding papers of this series). The character of geodesic dynamics is very probably related to the curvature properties of the host space-time (and its submanifolds to which the motion is restricted), though any ``generic" attempt to ascribe such global features of motion to the local space-time geometry deserves much standoff (see e.g. \cite{VieiraL-96b}). On the other hand, the complete geodesic integrability is known to be connected with the existence of the Killing--Yano tensors which in turn appears to be restricted to only some space-time curvature types (Petrov type D) (see e.g. \cite{Batista-15} and references therein). This suggests where a more specific connection between curvature and chaos could be found.

\begin{acknowledgments}
We are grateful for support from the grants GAUK-369015 and SVV-260211 of the Charles University (M.B.), and GACR-14-37086G of the Czech Science Foundation (O.S.).
O.S. also thanks M. Crosta for hospitality at Osservatorio Astrofisico di Torino and T. Ledvinka for advice on {\sc MAPLE}.
\end{acknowledgments}

\appendix

\section{Null geodesics tangent to the horizon}
\label{appendix-A}

Here the claim is proven that the curves $r=M\left[1\pm\cos(\theta-\theta_0)\right]$ of equation (\ref{null-geodesics}), along which we extended the metric inside the black hole, represent null geodesics starting tangentially from the horizon $r=2M$.
First, substituting ${\rm d}r=\mp M\sin(\theta-\theta_0)\,{\rm d}\theta$ into the $(r,\theta)$-part of the metric (\ref{metric}), one can check directly
\begin{align}
  &{\rm d}s^2(t\!=\!{\rm const},\phi\!=\!{\rm const})= \nonumber \\
  &= e^{2\lambda_{\rm ext}-2\nu_{\rm ext}}
     \left(\frac{{\rm d}r^2}{1-\frac{2M}{r}}+r^2{\rm d}\theta^2\right) \nonumber \\
  &= M^2\,\frac{e^{2\lambda_{\rm ext}}}{e^{2\nu_{\rm ext}}}
     \left\{\!\frac{\sin^2(\theta\!-\!\theta_0)}{1\!-\!\frac{2}{1\pm\cos(\theta-\theta_0)}}
           +[1\pm\cos(\theta\!-\!\theta_0)]^2\!\right\}
     {\rm d}\theta^2 \nonumber \\
  &= 0 \;.  \label{ds^2=0}
\end{align}
Even more immediately it is seen from the metric (\ref{metric,theta12}), given that the world-lines in question correspond to constant $\theta_-$ or $\theta_+$.

Second, let us verify whether such curves follow from a generic formula for photon geodesics. For a {\em Schwarzschild} field, such a formula was found by \cite{Darwin-59}; considering the ``scattering-type" motion in the $(r,\theta)$ plane ($\phi={\rm const}$) and parametrizing it so that its $\theta$ increases from a turning point at $(r=r_0,\theta=0)$, the formula reads
\begin{align}
  \theta(r)
  &= \frac{2\,\sqrt{r_0}\;\,[K(k)-F(\chi,k)]}{[(r_0-2M)(r_0+6M)]^{1/4}} \nonumber \\
  &= \frac{2\,\sqrt{r_0}\;\,F(\chi',k)}{[(r_0-2M)(r_0+6M)]^{1/4}} \;,  \label{Darwin}
\end{align}
where $F(\chi,k):=\int_0^\chi\frac{{\rm d}\alpha}{\sqrt{1-k^2\sin^2\alpha}}$
is the elliptic integral of the 1st kind, with amplitudes $\chi$, $\chi'$ and modulus $k$ given by
\begin{align*}
  &\sin^2\chi := 1-\frac{2M}{k^2}\,
                 \frac{1-\frac{r_0}{r}}{\sqrt{(r_0-2M)(r_0+6M)}} \;, \\
  &\sin^2\chi' := \frac{1-\sin^2\chi}{1-k^2\sin^2\chi}= \nonumber \\
       & \quad  = \frac{4M}{k^2}
                  \frac{1-\frac{r_0}{r}}
                       {\sqrt{(r_0-2M)(r_0+6M)}+r_0-2M-4M\frac{r_0}{r}} \;, \\
  &2k^2 := 1-\frac{r_0-6M}{\sqrt{(r_0-2M)(r_0+6M)}} \;,
\end{align*}
and $K(k):=F(\pi/2,k)$ is its complete version.
The ``scattering" orbits correspond to {\em pericentre} at $r_0>3M$, but the formula is also applicable for {\em apocentre} at $r_0<3M$; however, since $k^2>1$ in that case, one has to first transform to reciprocal modulus for such trajectories,
\[F(\chi',k)=\frac{1}{k}\;F(\psi,1/k),
  \qquad {\rm where} \quad \sin^2\psi=k^2\sin^2\chi',\]
which leads to the form
\begin{align*}
  &\theta(r)=
  \frac{2\,\sqrt{2r_0}\;\,
        F\!\left(\psi,\frac{\sqrt{2}\;[(r_0-2M)(r_0+6M)]^{1/4}}
                         {\left[\sqrt{(r_0-2M)(r_0+6M)}\,+6M-r_0\right]^{1/2}}\right)}
       {\left[\sqrt{(r_0-2M)(r_0+6M)}\,+6M-r_0\right]^{1/2}} \;, \\
  &\sin^2\psi = \frac{4M\left(1-\frac{r_0}{r}\right)}
                     {\sqrt{(r_0-2M)(r_0+6M)}+r_0-2M-4M\frac{r_0}{r}} \;.
\end{align*}
In the limit $r_0\rightarrow 2M$, this yields
\begin{align*}
  \theta(r)
   &= 2\,F\!\left(\!\arccos\sqrt{\frac{r}{2M}}\,,\;
                 \sqrt{2}\left(\frac{r_0}{2M}-1\right)^{1/4}\!\!\rightarrow 0\right)= \\
   &= 2\,\arccos\sqrt{\frac{r}{2M}} \\
  \Longleftrightarrow &\quad \cos^2\frac{\theta}{2}=\frac{r}{2M}
   \quad \Longleftrightarrow \quad r=M(1+\cos\theta) \;.
\end{align*}
Hence, in the {\em Schwarzschild} space-time, a photon starting tangentially to the horizon from $\theta=0$ really follows the curve $r=M(1+\cos\theta)$ and hits the singularity at $\theta=\pi$.

Importantly, it is in fact superfluous to speak of {\em projection} of the curves onto the $(r,\theta)$-plane, because their other components are zero anyway: (i) the $\phi$-motion is simply zero by assumption (and stays such due to axial symmetry); and (ii) the $t$-velocity also vanishes, because anything with finite locally measured energy $\hat{E}$ at the horizon has zero energy with respect to infinity, $E=\sqrt{-g_{tt}}\,\hat{E}$, but {\em for geodesics} the latter is {\em conserved} in stationary fields, which implies $u^t=0$ elsewhere thanks to the relation $E:=-u_t=-g_{tt}u^t$.
Hence, the relation $r=M\left[1\pm\cos(\theta-\theta_0)\right]$ actually provides {\em complete} information about this limit case of orbits having ``apocentre" at $r=2M$.

We have been having pure Schwarzschild in mind up to now, but if external sources are present, this clearly remains true, only $-g_{tt}$ changes from $|1-2M/r|$ to $|1-2M/r|\,e^{2\nu_{\rm ext}}$ (with $\nu_{\rm ext}$ finite on the horizon) -- see the metric (\ref{metric}).
This metric also implies that if the trajectory has no time and azimuthal components, the null result (\ref{ds^2=0}) holds regardless of the external sources, because the meridional interval is just scaled by $e^{2\lambda_{\rm ext}-2\nu_{\rm ext}}$ (which is again finite).
Let us stress that the above does not mean that the external sources have no effect, it just reflects that the Schwarzschild coordinates are well adapted to this effect (in particular, they keep the horizon on $r=2M$ regardless of the sources).

%%%%%%%%%%%%%%%%%%%%%%%%%%%%%%%%%%%%%%%%%%%%%%%%%%%%%%%%%%%%%%%%%
\begin{figure}[h!]
\centering
\includegraphics[width=\columnwidth]{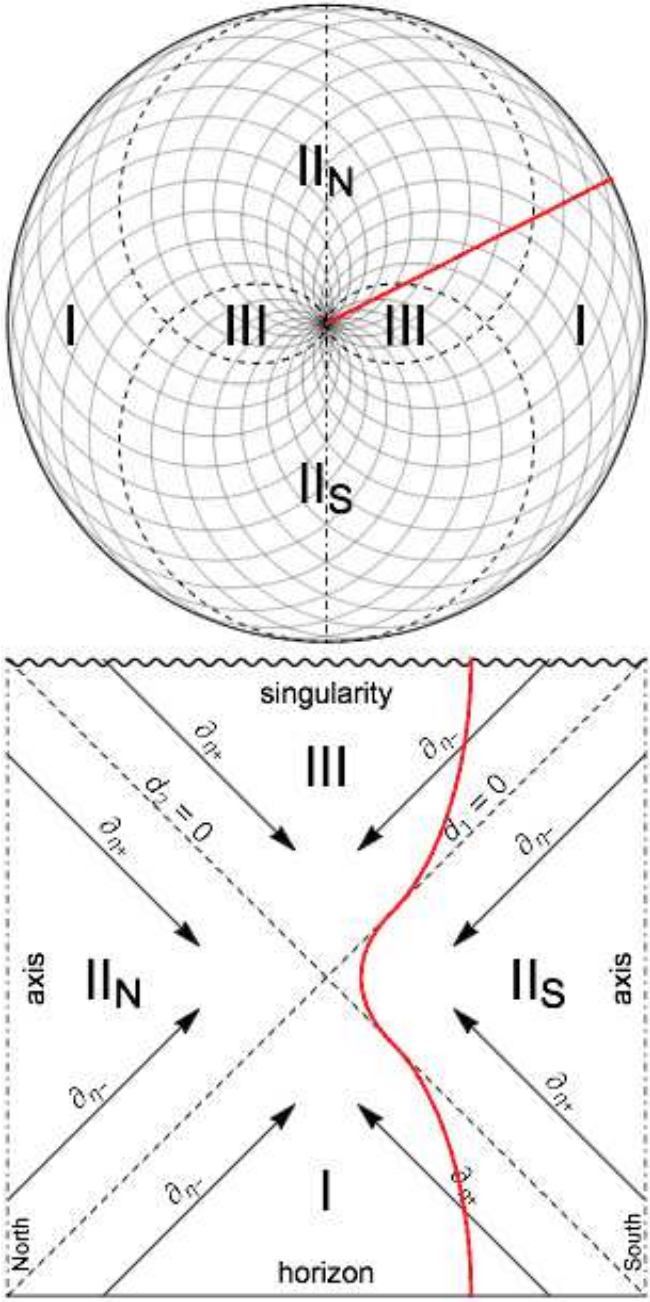}
\caption
{The four below-horizon regions with different combinations of $d_1$ and $d_2$ signs: I (where $d_1>0$, $d_2>0$), II$_{\rm N}$ (where $d_1<0$, $d_2>0$), II$_{\rm S}$ (where $d_1>0$, $d_2<0$) and III (where $d_1<0$, $d_2<0$).
At the {\em top} plot, they are shown in the Schwarzschild plot endowed with null geodesics descending from the horizon. At the {\em bottom} plot, they are shown as four triangles composed into a square. The domain I is confined by the horizon (solid line), the domains II$_{\rm N}$ and II$_{\rm S}$ contain respective parts of the symmetry axis (dot-dashed lines) and the domain III has the singularity (toothed line) at its inner rim. The arrows indicate directions of $\eta_{\pm}$ in each domain. The red solid line is included in order to show how the Schwarzschild radial direction is transformed.}
\label{inside-BH-diagram}
\end{figure}
%%%%%%%%%%%%%%%%%%%%%%%%%%%%%%%%%%%%%%%%%%%%%%%%%%%%%%%%%%%%%%%%%

\section{Weyl coordinates inside a black hole?}
\label{appendix-B}

Extension of the Weyl coordinates inside the black hole is somewhat problematic, because their relation to Schwarzschild coordinates is not bijective there. According to equations (\ref{Weyl-Schw}), modified to
\[\varrho=\sqrt{r(2M-r)}\,\sin\theta, \quad z=(r-M)\cos\theta,\]
any Schwarzschild-coordinate location $(r\!<\!2M,\theta)$ is readily and uniquely translated to Weyl coordinates, but the inverse transformation (\ref{Schw-Weyl}) has to be treated more carefully. Namely, the relations
\begin{align*}
  d_{1,2}&=\sqrt{(z\mp M)^2-\varrho^2}
          =\sqrt{(z\mp M+\varrho)(z\mp M-\varrho)} \\
         &=r-M\mp M\cos\theta
\end{align*}
clearly cannot be used everywhere, because the Schwarzschild-coordinate expressions for $d_{1,2}$ change sign on the curves $r=M(1\pm\cos\theta)$ (these were discussed in previous Appendix -- see the blue lines in figure \ref{null-geodesics-inside}, in particular). The inverse transformation
\[r-M=\frac{d_2+d_1}{2} \,, \qquad M\cos\theta=\frac{d_2-d_1}{2}\]
can still be used, but only after supplying correct signs to $d_{1,2}$ (whose Weyl forms are represented by square roots, providing just absolute values). In figure \ref{inside-BH-diagram} a diagram is shown of four different regions inside the black hole where the $d_{1,2}$ ``distances" have different combinations of signs; see a similar diagram in \cite{FrolovS-07} (figure 1). Note that in order to cover the whole black-hole interior, one has to supply all the four sign combinations ``by hand", they cannot be inferred from the Weyl-coordinate expressions of $d_{1,2}$ themselves.

Let us add that the problem of extension inside the static axisymmetric black hole has also been tackled, besides \cite{FrolovS-07} and in a slightly different manner, by \cite{PilkingtonMFB-11} when studying the character of a horizon of Schwarzschild-type space-times deformed by a set of multipoles. It was found there that the isolated horizon is always present, but need not represent a future outer trapping horizon nor a marginally trapped surface any more. However, as also warned in that paper, the distortions considered there were rather extreme (typically not asymptotically flat and often even violating the strong energy condition).

\bibliography{deformed-BHs-2.bib}

\end{document}